\documentclass[11pt]{article}
\usepackage{axodraw}
\usepackage{epsfig}
\usepackage{amsfonts}
\usepackage{amsmath}
\usepackage{bbm}
\input pix.sty

\newcommand{\sumint}[1]{\mbox{$\sum$}\!\!\!\!\!\!\!\int_{#1}}
\renewcommand{\eq}{eq.~}
\renewcommand{\eqs}{eqs.~}
\renewcommand{\se}{sec.~}

\renewcommand{\fig}{fig.~}
\renewcommand{\figs}{figs.~}
\newcommand{\dd}{\mathrm{d}}
\newcommand{\tinymsbar}{{\overline{\mbox{\tiny\rm{MS}}}}}
\newcommand{\Lambdamsbar}{{\Lambda_\tinymsbar}}

\newcommand{\Nc}{N_{\rm c}}

\newcommand{\Tc}{T_{\rm c}}
\newcommand{\gB}{g_\rmii{B}}
\newcommand{\mE}{m_\rmii{E}}

\newcommand{\gammaE}{\gamma_\rmii{E}}

\newcommand{\rmO}{{\mathcal{O}}}
\newcommand{\bmu}{\bar\Lambda} %{\bar\mu}

\def\lsi{\raise0.3ex\hbox{$<$\kern-0.75em\raise-1.1ex\hbox{$\sim$}}}
\def\gsi{\raise0.3ex\hbox{$>$\kern-0.75em\raise-1.1ex\hbox{$\sim$}}}
\newcommand{\lsim}{\mathop{\lsi}}
\newcommand{\gsim}{\mathop{\gsi}}

\newcommand{\rmii}[1]{{\mbox{\tiny\rm{#1}}}}

\newcommand{\Tint}[1]{{\hbox{$\sum$}\!\!\!\!\!\!\!\int\,}_{\!\!\!\!\raise-0.9ex\hbox{$\scriptstyle{#1}$}}}
\newcommand{\Tinti}[1]{{{\Sigma}\!\!\!\!\raise0.3ex\hbox{$\int$}_\rmii{${#1}$}}}
\renewcommand{\Tint}[1]{\sumint{#1}}

\newcommand{\bi}{\begin{itemize}}
\newcommand{\ei}{\end{itemize}}
\newcommand{\hide}[1]{ }

\def\TAsc(#1,#2)(#3,#4,#5)%
{\SetWidth{2.0}\CArc(#1,#2)(#3,#4,#5)\SetWidth{1.0}}
\def\Lwidth{3}

\def\TAgl(#1,#2)(#3,#4,#5){\SetWidth{2.0}\PhotonArc(#1,#2)(#3,#4,#5){\Lwidth}%
{6.283 #3 mul 360 div #4 #5 sub #4 #5 sub mul sqrt mul Tdensity mul}%
\SetWidth{1.0}}
\def\TLgl(#1,#2)(#3,#4){\SetWidth{2.0}\Photon(#1,#2)(#3,#4){\Lwidth}
{#1 #3 sub #1 #3 sub mul #2 #4 sub #2 #4 sub mul add sqrt Tdensity mul}%
\SetWidth{1.0}}
\renewcommand{\picc}[1]{\;\parbox[c]{60pt}{\begin{picture}(60,30)(0,0)
\SetWidth{1.0}\SetScale{1.3} #1 \end{picture}}\; }
\def\Lwidth{1.3}

\newcommand{\picu}[1]{\;\parbox[c]{60pt}{\begin{picture}(60,30)(0,0)
\SetWidth{1.0}\SetScale{1.0} #1 \end{picture}}\; }
\def\EleA{\picu{%
 \Agl(30,5)(22.3,27,153)%
 \Agl(30,25)(22.3,207,333)%
 \COval(10,15)(2,2)(0){Black}{Black}%
 \COval(50,15)(2,2)(0){Black}{Black}%
}}
\def\EleB{\picu{%
 \Agl(30,5)(22.3,27,153)%
 \Agl(30,25)(22.3,207,333)%
 \COval(10,15)(2,2)(0){Black}{Black}%
 \COval(50,15)(2,2)(0){Black}{Black}%
 \Agl(58,15)(8,0,360)%
}}
\def\EleC{\picu{%
 \Agl(30,5)(22.3,27,153)%
 \Agl(30,25)(22.3,207,333)%
 \COval(10,15)(2,2)(0){Black}{Black}%
 \COval(50,15)(2,2)(0){Black}{Black}%
 \Lgl(10,15)(50,15)%
}}
\def\EleD{\picu{%
 \Agl(30,5)(22.3,90,153)%
 \Agl(30,25)(22.3,207,333)%
 \COval(10,15)(2,2)(0){Black}{Black}%
 \COval(50,15)(2,2)(0){Black}{Black}%
 \Agl(43,27)(12,180,300)%
 \Agl(38,16)(12,0,120)%
}}
\def\EleE{\picu{%
 \Agl(30,5)(22.3,27,153)%
 \Agl(30,25)(22.3,207,333)%
 \COval(10,15)(2,2)(0){Black}{Black}%
 \COval(50,15)(2,2)(0){Black}{Black}%
 \GCirc(30,27.3){4}{0.5}
}}
\def\EleF{\picu{%
 \Agl(20,10)(11.15,27,153)%
 \Agl(20,20)(11.15,207,333)%
 \Agl(40,10)(11.15,27,153)%
 \Agl(40,20)(11.15,207,333)%
 \COval(10,15)(2,2)(0){Black}{Black}%
 \COval(50,15)(2,2)(0){Black}{Black}%
}}
\def\EleG{\picu{%
 \Agl(30,5)(22.3,27,153)%
 \Agl(30,25)(22.3,207,333)%
 \COval(10,15)(2,2)(0){Black}{Black}%
 \COval(50,15)(2,2)(0){Black}{Black}%
 \Lgl(30,2.7)(30,27.3)%
}}
\makeatletter \@addtoreset{equation}{section} \makeatother
\renewcommand{\theequation}{\arabic{section}.\arabic{equation}}
\makeatletter
\renewcommand\section{\@startsection {section}{1}{\z@}%
                                   {-5.5ex \@plus -1ex \@minus -.2ex}% bfr-
                                   {2.3ex \@plus.2ex}%
                                   {\normalfont\large\bfseries}}
\renewcommand\subsection{\@startsection{subsection}{2}{\z@}%
                                     {-3.25ex\@plus -1ex \@minus -.2ex}%
                                     {1.5ex \@plus .2ex}%
                                     {\normalfont\normalsize\bfseries}}
\renewcommand\thesection {\@arabic\c@section}
\renewcommand\thesubsection   {\thesection.\@arabic\c@subsection}
\renewcommand{\@seccntformat}[1]{%
\csname the#1\endcsname.\hspace{1.0em}}
\makeatother
\begin{document}

\flushbottom

\begin{titlepage}

\begin{flushright}
% OUTLINE  \\
% DRAFT \\
BI-TP 2010/42\\
% arXiv:yymm.nnnn\\
\vspace*{1cm}
\end{flushright}
\begin{centering}
\vfill

{\Large{\bf
Intermediate distance correlators in hot Yang-Mills theory
}}

\vspace{0.8cm}

M.~Laine$^{\rm a}$, %%\footnote{laine@physik.uni-bielefeld.de}
M.~Veps\"al\"ainen$^{\rm b}$, %%\footnote{mtvepsal@pcu.helsinki.fi}
A.~Vuorinen$^{\rm a}$ %%\footnote{vuorinen@physik.uni-bielefeld.de}

\vspace{0.8cm}

$^\rmi{a}$%
{\em
Faculty of Physics, University of Bielefeld,
D-33501 Bielefeld, Germany\\}

\vspace{0.3cm}

$^{\rm b}$%
{\em
Department of Physics,
P.O.Box 64, FI-00014 University of Helsinki, Finland\\}

\vspace*{0.8cm}

\mbox{\bf Abstract}

\end{centering}

\vspace*{0.3cm}

\noindent
Lattice measurements of spatial correlation functions of the
operators $FF$ and $F\tilde F$ in thermal SU(3) gauge theory 
have revealed a clear difference between the two channels at
``intermediate'' distances, $x \sim 1/\pi T$.  This is at odds with
the AdS/CFT limit which predicts the results to coincide.  On the
other hand, an OPE analysis at short distances ($x \ll 1/\pi T$) as
well as effective theory methods at long distances ($x \gg
1/\pi T$) suggest differences. Here we study 
the situation at intermediate distances by determining 
the time-averaged spatial correlators through a 2-loop 
computation. We do find unequal results, however the numerical 
disparity is small. Apart from theoretical issues, 
a future comparison of our results with time-averaged lattice 
measurements might also be of phenomenological interest in 
that understanding the convergence of the weak-coupling series 
at intermediate distances may bear on studies of the thermal 
broadening of heavy quarkonium resonances.

\vfill

%% %\noindent
%% %PACS numbers:
%% %11.10.Wx, %        Finite temperature field theory
%% { %11.15.Ha, %        Lattice gauge theory }
%% %12.38.Bx, %        Perturbative calculations in QCD
%% %12.38.Mh, %        Quark--gluon plasma
%% %14.40.Nd, %        Bottom mesons
%% %\\
%% %Keywords:

\vspace*{1cm}

\noindent
November 2010

\vfill

\end{titlepage}

%%%%%%%%%%%%%%%%%%%%%%%%%%% SECTION %%%%%%%%%%%%%%%%%%%%%%%%%%%%%%%%%%%%%%
%
\section{Introduction}

Spatial correlation functions of local gauge-invariant operators
(such as components of the energy-momentum tensor or various charge 
densities) offer a set of observables that can yield a detailed 
characterization of the physical properties of a thermal system. 
Consequently, in lattice QCD, correlators related to gluonic 
(flavour-singlet) as well as hadronic (flavour non-singlet) operators 
have been studied since dawn~\cite{DeGrand:1986uf,DeTar:1987xb} 
and continue to be important in the present era of unquenched
simulations (see, e.g.,\ refs.~\cite{Maezawa:2010vj,Cheng:2010fe}).

At high temperatures ($T \gg 200$~MeV), 
spatial correlation functions can be sensitive to 
different types of physics at different distance scales. At short
distances ($x \ll 1/\pi T$), we expect them to display vacuum-like 
behaviour; at intermediate distances ($x \sim 1/\pi T$), we
expect a complicated functional form with thermal modifications 
of order unity; at long distances ($x \gg 1/\pi T$), the 
correlators should ultimately reduce to a single exponential, 
characterized by collective phenomena such as colour-electric 
or colour-magnetic screening~\cite{Arnold:1995bh}.
In addition, the zero-temperature confinement scale may 
also make an appearance, if a measurement at a low
temperature is subtracted from the high-temperature measurement
in an effort to remove the vacuum-like behaviour at short distances.

For future reference, 
we note that the distance scale that one is interested in may influence 
some technical details of the lattice measurement 
carried out. For instance, 
for short-distance physics a signal may be obtained from point-point 
correlators, whereas the long-distance asymptotics can perhaps 
best be extracted by using operators averaged over the Euclidean 
time direction as well as over the transverse spatial plane. 

In the present paper, following a recent high-precision 
lattice measurement~\cite{Iqbal:2009xz}, we consider spatial 
correlators within pure SU(3) gauge theory at a temperature
above that of the deconfinement transition. The operators
correlated are $\tr[F_{\mu\nu} F_{\mu\nu}]$
(corresponding physically to the trace of 
the energy-momentum tensor, modulo a coefficient)
and $\tr[F_{\mu\nu} \tilde F_{\mu\nu}]$
(corresponding physically to topological 
charge density, modulo a coefficient).
We assume the operators to be averaged over the Euclidean 
time coordinate but to remain point-like in space 
coordinates. On this point we actually differ from 
ref.~\cite{Iqbal:2009xz} in which the operators were 
point-like also in time; apart from technical 
simplifications, our choice is motivated by the fact that
the correlators then become more analogous to Polyakov-loop 
ones which are relevant for the phenomenology
of heavy quarkonium physics (the ramifications in this 
direction will be discussed in the conclusions). 
Time-averaged correlators are also the ones 
containing the thermodynamic information relevant
for spectral sum rules (see, e.g., refs.~\cite{rs,hbm_d}).

One of the main findings of ref.~\cite{Iqbal:2009xz} was that, 
once the (genuine) vacuum part was subtracted from 
the correlators, the two channels showed strikingly different 
behaviours at intermediate distances 
(cf.\ \fig5 of ref.~\cite{Iqbal:2009xz}):
for $\tr[F_{\mu\nu} F_{\mu\nu}]$ the thermal modification was
negative in this range (reducing the correlator) whereas 
for $\tr[F_{\mu\nu} \tilde F_{\mu\nu}]$ it was positive 
(enhancing the correlator). That such a difference exists
is in contrast with the strongly-coupled AdS/CFT limit of $\mathcal{N} = 4$
Super-Yang-Mills theory which 
predicts the two correlators to coincide~\cite{Son:2002sd,Iqbal:2009xz}. 

Within pure SU(3) theory, the emergence of 
a qualitative difference between the two channels has previously  
been understood both in the short-distance ($x\ll 1/\pi T$)
and in the long-distance ($x\gg 1/\pi T$) limits. In the 
short-distance limit, the correlators can be expanded in 
an Operator Product Expansion (OPE). The leading terms are
vacuum-like contributions, which already differ as far as 
non-logarithmic terms are concerned~\cite{old}. Furthermore, 
OPE contributions proportional to the expectation 
value of the trace anomaly, $(e-3p)(T)$, come with opposite 
signs~\cite{ope,lvv}, underlining that any breaking
of conformal invariance is bound to lift the degeneracy.  

On the long-distance side ($x\gg 1/\pi T$), the very fact
that the two operators have different discrete quantum 
numbers has been exploited for {\em defining} 
colour-electric and colour-magnetic screening masses 
in a non-Abelian theory~\cite{Arnold:1995bh}. The colour-electric
channel couples to $\tr[F_{\mu\nu} \tilde F_{\mu\nu}]$
and the colour-magnetic to $\tr[F_{\mu\nu} F_{\mu\nu}]$. Practical 
measurements in pure Yang-Mills theory and QCD (making use 
of an infrared effective field theory) can be found in 
ref.~\cite{Hart:2000ha} and indeed show substantial variation
in the screening masses; the qualitative differences 
in the corresponding weak-coupling expressions 
have been elaborated upon in ref.~\cite{Laine:2009dh}.

The purpose of the present paper is to ``interpolate'' between
the known short-distance and long-distance limits, by considering
intermediate distances $x\sim 1/\pi T$. This situation is ``simple''
in the sense that straightforward perturbation theory is formally
applicable, without the need of re-organization in terms of OPE
or of effective field theories. The challenge is that the correlators
have a relatively complicated functional form. Nevertheless, 
we have managed to obtain expressions for both correlators, 
and foresee an opportunity for a practical comparison of 
the weak-coupling expansion against lattice simulations of 
the type in ref.~\cite{Iqbal:2009xz}. 

The plan of this paper is the following. 
The observables considered are specified in \se\ref{se:setup}, 
and the main steps of the computation are described in some 
detail in \se\ref{se:naive}. Section~\ref{se:results} collects 
together the basic results applicable for short and 
intermediate distances, whereas in \se\ref{se:ir} we elaborate 
on how the perturbative expressions need to be resummed in 
the long-distance limit. Numerical illustrations 
comprise \se\ref{se:plot}, and our conclusions are collected
in \se\ref{se:concl}.

%%%%%%%%%%%%%%%%%%%%%%%%%%%% SECTION %%%%%%%%%%%%%%%%%%%%%%%%%%%%%%%%%%%%
%
\section{Setup}
\la{se:setup}

Our basic notation follows ref.~\cite{lvv}, so we 
discuss the setup only briefly. We consider pure SU($\Nc$)
Yang-Mills theory (with $\Nc = 3$ in numerical estimates), 
dimensionally regularized by analytically continuing 
to $D=4-2\epsilon$ space-time dimensions
($S_\rmi{E} = \int_0^{\beta} \! {\rm d}\tau 
\int \! {\rm d}^{3-2\epsilon}\vec{x}\, \fr14 F^a_{\mu\nu} F^a_{\mu\nu}$). 
The operators considered are 
\be
 \theta \equiv c_\theta\, \gB^2 F^a_{\mu\nu}F^a_{\mu\nu}
 \;, \quad
 \chi \equiv c_\chi\, \epsilon_{\mu\nu\rho\sigma} 
 \gB^2 F^a_{\mu\nu}F^a_{\rho\sigma}
 \;, \la{ops}
\ee
where $\gB^2$ is the bare gauge coupling squared; these operators
require no renormalization at the order of our computation.
We do not need to specify the coefficients $c_\theta$, $c_\chi$, 
but note that in ref.~\cite{Iqbal:2009xz} the values
$
  c_\theta \approx  
 -\frac{b_0}{2} 
 \;, 
 c_\chi \equiv \frac{1}{64\pi^2} 
$
were chosen, where $b_0 \equiv \frac{11\Nc}{3(4\pi)^2}$. 
The bare gauge coupling squared can be expanded in terms of 
the renormalized one, $g^2$, as 
\be
 \gB^2 = g^2 \Lambda^{2\epsilon} \biggl[
 1 - \frac{11 g^2 \Nc}{3(4\pi)^2} \frac{1}{\epsilon} + \ldots 
 \biggr] 
 \;, \la{gB}
\ee
where $\Lambda$ denotes a scale parameter. The 
$\msbar$ scheme renormalization scale is denoted by 
$ 
 \bmu^{2} \equiv {4\pi} \Lambda^2  
 {e^{-\gammaE}}
$.
In order to avoid unnecessary clutter
all appearances of $\Lambda^{2\epsilon}$, which cancel 
when the result is expressed in terms of the 
renormalized gauge coupling, will be suppressed. 

We wish to evaluate the behaviour of the  
correlators of $\theta$ and $\chi$ as functions 
of the spatial distance $x\equiv |\mathbf{x}|$, integrated 
over the Euclidean time $\tau$. To this end, we define 
the quantities 
($X\equiv (\tau,\mathbf{x})$; 
$P\equiv (p_n,\mathbf{p})$; 
% $p \equiv |\mathbf{p}|$; 
$G_\theta(X) \equiv  
 \langle \theta(X) \theta(0) \rangle_\rmi{c}$; 
$G_\chi(X) \equiv 
 \langle \chi(X) \chi(0) \rangle$)
\ba
 \bar{G}_\theta(x)\equiv \int_0^\beta {\rm d}\tau\, G_\theta(X)
 \;, \quad
 \bar{G}_\chi(x)\equiv \int_0^\beta {\rm d}\tau\, G_\chi(X)
 \; .
\ea
Using the corresponding Fourier transforms
\be
 \tilde G_\theta(P) \equiv \int_X e^{- i P\cdot X} G_\theta(X)
 \;, \quad
 \tilde G_\chi(P) \equiv \int_X e^{- i P\cdot X} G_\chi(X)
 \;, \la{GP}
\ee
we see that for our purposes it suffices to consider 
the static ($p_n=0$) limit of the correlators in momentum space:
\ba
 \bar{G}_\theta(x) = \int_\vec{p} e^{i \mathbf{p}\cdot \mathbf{x}}
 \, 
 \tilde G_\theta(p_n=0,\mathbf{p})
 \;,\quad
 \bar{G}_\chi(x) = \int_\vec{p} e^{i \mathbf{p}\cdot \mathbf{x}}\, 
 \tilde G_\chi(p_n=0,\mathbf{p})
 \;. \la{pn0}
\ea

We work up to next-to-leading order (NLO) in perturbation theory. 
A natural starting point for the computation are then 
the expressions in \eqs(3.1) and (3.2) of ref.~\cite{lvv}, 
obtained after ``scalarizing'' the 1-loop and 2-loop Feynman 
graphs (\fig\ref{fig:graphs}) contributing to the 
momentum-space correlators ($d_A \equiv \Nc^2-1$):
\ba
 && \hspace*{-1cm} \frac{\tilde G_\theta(P)}{4 d_A c_\theta^2 \gB^4} =
% \nn & = &
 (D-2) \biggl[ -\mathcal{J}_\rmi{a} + \fr12 \mathcal{J}_\rmi{b} \biggr]
 \nn
 & + & \gB^2 \Nc \biggl\{
 2 (D-2) \biggl[ - (D-1)\mathcal{I}_\rmi{a} + (D-4) \mathcal{I}_\rmi{b} \biggr]
 + (D-2)^2 \biggl[ \mathcal{I}_\rmi{c} - \mathcal{I}_\rmi{d} \biggr]
 \nn & & \quad + \,
 \frac{22-7D}{3} \mathcal{I}_\rmi{f} - \frac{(D-4)^2}{2} \mathcal{I}_\rmi{g}
 + (D-2)
 \biggl[
   -3 \mathcal{I}_\rmi{e} + 3 \mathcal{I}_\rmi{h} + 2\mathcal{I}_\rmi{i}
    - \mathcal{I}_\rmi{j}
 \biggr] \biggr\} % + \rmO(\gB^4)
 \;, \la{Gtheta_bare} \\
%%%%%%%%%%%%%%%%%%%%%%%%%%%%%%%%%%%%%%
%%%%%%%%%%%%%%%%%%%%%%%%%%%%%%%%%%%%%%
 && \hspace*{-1cm} \frac{\tilde G_\chi(P)}{-16 d_A c_\chi^2 \gB^4 (D-3)} =
% \nn & = &
 (D-2) \biggl[ -\mathcal{J}_\rmi{a} + \fr12 \mathcal{J}_\rmi{b} \biggr]
 \nn
 & + & \gB^2 \Nc \biggl\{
 2 (D-2) \biggl[ - \mathcal{I}_\rmi{a} + (D-4) \mathcal{I}_\rmi{b} \biggr]
 + (D-2)^2 \biggl[ \mathcal{I}_\rmi{c} - \mathcal{I}_\rmi{d} \biggr]
 \nn & & \quad - \,
 \frac{2 D^2-17D+42}{3} \mathcal{I}_\rmi{f} - 2 (D-4) \mathcal{I}_\rmi{g}
 + (D-2)
 \biggl[
   -3 \mathcal{I}_\rmi{e} + 3 \mathcal{I}_\rmi{h} + 2\mathcal{I}_\rmi{i}
    - \mathcal{I}_\rmi{j}
 \biggr] \biggr\} % + \rmO(\gB^4)
 \;. \la{Gchi_bare}
\ea
The definitions of the master sum-integrals 
$\mathcal{J}_\rmi{a} \ldots \mathcal{I}_\rmi{j}$
can be found in \eqs\nr{Ja}--\nr{Ij} below. 

It will become clear later on that at long distances, 
the NLO corrections can overtake the leading order (LO) 
terms, indicating a breakdown of the perturbative series. 
In this situation perturbation theory needs to be resummed
through effective field theory techniques. We return to 
when this happens 
in \se\ref{se:ir}; for the moment we simply compute 
the correlators as they stand in \eqs\nr{pn0}--\nr{Gchi_bare}. 
The results will be denoted 
by $\bar G^\rmi{F}_{\theta,\chi}$, 
with ``$\rmi{F}$'' standing for ``full theory''. 

%%%%%%%%%%%%%%%%%%%%%%%%% FIGURE %%%%%%%%%%%%%%%%%%%%%%%%%%%%%%%%%%%%%%%%%
%
\begin{figure}[t]

\hspace*{1.5cm}%
\begin{minipage}[c]{3cm}
\begin{eqnarray*}
&&
 \hspace*{-1cm}
 \EleA
\\[1mm]
&&
 \hspace*{0.0cm}
 \mbox{(i)}
\end{eqnarray*}
\end{minipage}%
\begin{minipage}[c]{10cm}
\begin{eqnarray*}
&&
 \hspace*{-1cm}
 \EleB \quad\;
 \EleC \quad\;
 \EleD \quad\;
\\[1mm]
&&
 \hspace*{0.0cm}
 \mbox{(ii)} \hspace*{2.2cm}
 \mbox{(iii)} \hspace*{2.2cm}
 \mbox{(iv)}
\\[5mm]
&&
 \hspace*{-1cm}
 \EleE \quad\;
 \EleF \quad\;
 \EleG \quad
\\[1mm]
&&
 \hspace*{0.0cm}
 \mbox{(v)} \hspace*{2.2cm}
 \mbox{(vi)} \hspace*{2.2cm}
 \mbox{(vii)}
 \\[3mm]
\end{eqnarray*}
\end{minipage}

\caption[a]{\small
The LO and NLO Feynman graphs contributing to the correlators of 
\eqs\nr{Gtheta_bare}, \nr{Gchi_bare}.}
\la{fig:graphs}
\end{figure}
%
%%%%%%%%%%%%%%%%%%%%%%%%%%%%%%%%%%%%%%%%%%%%%%%%%%%%%%%%%%%%%%%%%%%%%%%%%%

%%%%%%%%%%%%%%%%%%%%%%%%%%% SECTION %%%%%%%%%%%%%%%%%%%%%%%%%%%%%%%%%%
%
\section{Naive full theory computation}
\la{se:naive}

In this section, we describe in some
detail the evaluation of the various master 
sum-integrals appearing in \eqs\nr{Gtheta_bare}, \nr{Gchi_bare}, 
once they are Fourier transformed according to \eq\nr{pn0}. 

%%%%%%%%%%%%%%%%%%%%% SUBSECTION %%%%%%%%%%%%%%%%%%%%%%%%%%%%%%%%%%%
%
\subsection{Sum-integrals \label{sec:ints}}

The master sum-integrals are defined using 
dimensional regularization in $D=4-2\epsilon$ dimensions 
and, as in ref.~\cite{lvv}, read
($\Tinti{Q} \equiv T \sum_{q_n} \int_\vec{q}$)
\ba
 \mathcal{J}_\rmi{a} & \equiv &
 \Tint{Q} \frac{P^2}{Q^2}
 \;, \la{Ja} \\
 \mathcal{J}_\rmi{b} & \equiv &
 \Tint{Q} \frac{P^4}{Q^2(Q-P)^2}
 \;, \la{Jb} \\
 \mathcal{I}_\rmi{a} & \equiv &
 \Tint{Q,R} \frac{1}{Q^2R^2}
 \;,\\
 \mathcal{I}_\rmi{b} & \equiv &
 \Tint{Q,R} \frac{P^2}{Q^2R^2(R-P)^2}
 \;,
\ea
\ba 
 \mathcal{I}_\rmi{c} & \equiv &
 \Tint{Q,R} \frac{P^2}{Q^2R^4}
 \;, \\
 \mathcal{I}_\rmi{d} & \equiv &
 \Tint{Q,R} \frac{P^4}{Q^2R^4(R-P)^2}
 \;, \\
 \mathcal{I}_\rmi{e} & \equiv &
 \Tint{Q,R} \frac{P^2}{Q^2R^2(Q-R)^2}
 \;, \\
 \mathcal{I}_\rmi{f} & \equiv &
 \Tint{Q,R} \frac{P^2}{Q^2(Q-R)^2(R-P)^2}
 \;, \\
 \mathcal{I}_\rmi{g} & \equiv &
 \Tint{Q,R} \frac{P^4}{Q^2(Q-P)^2R^2(R-P)^2}
 \;, \\
 \mathcal{I}_\rmi{h} & \equiv &
 \Tint{Q,R} \frac{P^4}{Q^2R^2(Q-R)^2(R-P)^2}
 \;, \la{def_Ih} \\
 \mathcal{I}_\rmi{i} & \equiv &
 \Tint{Q,R} \frac{(Q-P)^4}{Q^2R^2(Q-R)^2(R-P)^2}
 \;,\\
 \mathcal{I}_\rmi{i'} & \equiv &
 \Tint{Q,R} \frac{4(Q\cdot P)^2}{Q^2R^2(Q-R)^2(R-P)^2}
 \;, \\
 \mathcal{I}_\rmi{j} & \equiv &
 \Tint{Q,R} \frac{P^6}{Q^2R^2(Q-R)^2(Q-P)^2(R-P)^2}
 \;. \la{Ij}
\ea
As specified in \eq\nr{pn0} we are  interested in taking their 
three-dimensional (3d) Fourier transforms, \textit{i.e.~}the (dimensionally 
regularized) functions
\ba
 \bar{\mathcal{I}}_\rmi{m}(x) &\equiv& %\Lambda^{2\epsilon}
 \int_\vec{p} e^{i \mathbf{p}\cdot \mathbf{x}}
 \,\mathcal{I}_\rmi{m}(p_n=0,\vec{p})
 \;.
\ea
We adopt a notation in the following whereby $p$ stands for 
a four-vector with vanishing frequency, i.e.\ $p \equiv (0,\vec{p})$; 
in the above definitions, we can thus replace $P^2 \to p^2$,
$(Q-P)^2 \to (p-Q)^2$, and $(R-P)^2 \to (p-R)^2$.

In some cases, the contribution of the 
zero modes ($q_n=r_n=0$) to the master sum-integrals 
is delicate, and a proper handling requires the use 
of a consistent regularization procedure.  
We therefore separate the zero-mode contribution 
from the rest, and evaluate it 
in \se\ref{zeromode};
the results are denoted by
$\bar{\mathcal{I}}_\rmi{m}^0$.  
In subsequent sections, % \ref{se:IaIe}--\ref{se:Ij}, 
we assume 
that the zero modes have been subtracted from the sum-integrals;
the subtracted sum-intergals 
are denoted by $\bar{\mathcal{I}}_\rmi{m}'$.

%%%%%%%%%%%%%%%%%%%%%%%%%%%% SUBSECTION %%%%%%%%%%%%%%%%%%%%%%%%%%%%%%%
%
\subsection{Zero modes}
\label{zeromode}

Setting $q_n=r_n=0$, we see that many of the master sum-integrals 
vanish as scaleless integrals in dimensional regularization,
\ba
 \bar{\mathcal{J}}^0_\rmi{a}\;=\; 
 \bar{\mathcal{I}}^0_\rmi{a}\;=\; 
 \bar{\mathcal{I}}^0_\rmi{b}\;=\; 
 \bar{\mathcal{I}}^0_\rmi{c}\;=\; 
 \bar{\mathcal{I}}^0_\rmi{d}\;=\; 
 \bar{\mathcal{I}}^0_\rmi{e}\;=\;0
 \;. 
\ea
For the rest, we first note that by dimensional analysis,
\ba
 \mathcal{J}_\rmi{b}^0&\sim& T (p^2)^{\fr32-\epsilon}
 \;,\\
 \mathcal{I}^0_\rmi{m}&\sim& T^2 (p^2)^{1-2\epsilon}
 \;,
 \quad {\rm m} \in\,\{{\rm f,g,h,i,i',j }\}
 \;.
\ea
Recalling on the other hand that
\be
 \int_\vec{p} e^{i \mathbf{p}\cdot \mathbf{x}}
 \,p^{3-2\epsilon} = \fr{12}{\pi^2 x^6}
 +\rmO(\epsilon)
 \;, \quad
 \int_\vec{p} e^{i \mathbf{p}\cdot \mathbf{x}}
 \,p^{2-4\epsilon} = -\fr{6\epsilon}{\pi x^5}+\rmO(\epsilon^2)
 \;,
\ee
we observe that it suffices to evaluate the master integrals 
$\mathcal{I}^0_\rmi{m}$ to order $1/\epsilon$. An elementary calculation 
(making use of the ``zero-temperature parts'' of the various 
master sum-integrals given in appendix~A of ref.~\cite{lvv}, after
setting $D\to 3-2\epsilon$ in them) 
now produces
\ba
 \bar{\mathcal{J}}^0_\rmi{b}&=&
 \fr{96\pi^4T^7}{\bar{x}^6} +\rmO(\epsilon), \la{Jb0} \\
 \bar{\mathcal{I}}^0_\rmi{f} &=&
 -\fr{3\pi^2T^7}{\bar{x}^5} +\rmO(\epsilon), \la{If0} \\[2mm]
 \bar{\mathcal{I}}^0_\rmi{g} &=&\rmO(\epsilon),\\[1mm]
 \bar{\mathcal{I}}^0_\rmi{h} &=&
 \fr{3\pi^2T^7}{\bar{x}^5} +\rmO(\epsilon),\\
 \bar{\mathcal{I}}^0_\rmi{i} &=&
 \fr{3\pi^2T^7}{\bar{x}^5} +\rmO(\epsilon),\\[2mm]
 \bar{\mathcal{I}}^0_\rmi{i'} &=&\rmO(\epsilon),  \la{Iipzero} \\[1mm]
 \bar{\mathcal{I}}^0_\rmi{j} &=&
 \fr{6\pi^2T^7}{\bar{x}^5} +\rmO(\epsilon)
 \;,
\ea
where we have defined the dimensionless variable 
\be
 \bar{x}\equiv 2\pi T x
 \;, 
\ee
in terms of which we will present our results.

%%%%%%%%%%%%%%%%%%%%%%%%%%%% SUBSECTION %%%%%%%%%%%%%%%%%%%%%%%%%%%%%%%
%
\subsection{$\bar{\mathcal{J}}_\rmi{a}$ 
and $\bar{\mathcal{J}}_\rmi{b}$ \label{Js}}

The 1-loop sum-integrals $\bar{\mathcal{J}}_\rmi{a}$ 
and $\bar{\mathcal{J}}_\rmi{b}$ are only needed in the combination
\ba
 \bar{\mathcal{J}}_\rmi{b} - 2 \bar{\mathcal{J}}_\rmi{a} &=& 
 \int_\vec{p} e^{i\mathbf{p\cdot x}} p^2\,\Tint{Q}\fr{1}{Q^2}
 \bigg\{\fr{p^2}{(p-Q)^2}-2\bigg\}\nn
 & \rightarrow & 
 \int_\vec{p} e^{i\mathbf{p\cdot x}} \,\Tint{Q}\fr{p^4}{Q^2(p-Q)^2}
 \;,
\ea
where we have discarded a contact term. 
The remaining term can be evaluated in a straightforward 
manner by writing $p^4$ as a spatial derivative 
and using results derived in ref.~\cite{az}. 
We will later on also need the $\rmO(\epsilon)$ 
part but, for the time being, 
we merely write 
\ba
 \bar{\mathcal{J}}_\rmi{b} - 2 \bar{\mathcal{J}}_\rmi{a} 
 &=&  \fr{4\pi^4 T^7}{\bar{x}}\fr{{\rm d}^4}{{\rm d}\bar{x}^4}
 \biggl( \fr{{\rm coth}\bar{x}}{\bar{x}} \biggr) 
 + \rmO(\epsilon)
 \;. \la{resJb}
\ea
Note that this expression contains both the zero-mode 
and the non-zero mode contributions;
%($\bar{\mathcal{J}}_\rmi{b} = 
%\bar{\mathcal{J}}_\rmi{b}^0 + \bar{\mathcal{J}}_\rmi{b}'$);
the zero-mode contribution of \eq\nr{Jb0} is recovered as the 
only power-law term at long distances, if we expand
$\mathop{\rm coth}\bar x = 1 + \rmO(e^{-2\bar{x}})$.

%%%%%%%%%%%%%%%%%%%%%%%%%%%%%%%%%%%%%% SUBSECTION %%%%%%%%%%%%%%%%%%%%%
%
\subsection{$\bar{\mathcal{I}}_\rmi{a}$,
$\bar{\mathcal{I}}_\rmi{c}$,
$\bar{\mathcal{I}}_\rmi{e}$}
\la{se:IaIe}

In momentum space, the $P$-dependence of the sum-integrals 
$\mathcal{I}_\rmi{m}$, ${\rm m}\in\{{\rm a,c,e}\}$, factorizes from 
the $Q$ and $R$ dependences (in fact, $\mathcal{I}_\rmi{e}$ vanishes
identically). One is then left with taking a 3d Fourier 
transform of a non-negative integer power of the momentum squared, 
which leads to a result proportional to the delta function or its 
derivatives. The results could be directly extracted from ref.~\cite{lvv},
but in the present paper we are only interested in 
finite values $\bar{x}\sim 1$, 
so we set these integrals to zero,
\ba
 \bar{\mathcal{I}}_\rmi{m} \to 0
 \;,\quad 
 \rm{m}\in\{\rm{a,c,e}\}
 \;.
\ea
Contact terms also appear as parts of the other master sum-integrals, 
and will be omitted there as well; again, if needed, they can be 
extracted from ref.~\cite{lvv}.

%%%%%%%%%%%%%%%%%%%%%%%%% SUBSECTION %%%%%%%%%%%%%%%%%%%%%%%%%%%%%%%%
%
\subsection{$\bar{\mathcal{I}}_\rmi{b}$,
$\bar{\mathcal{I}}_\rmi{d}$,
$\bar{\mathcal{I}}_\rmi{g}$}

Next, consider the sum-integrals 
$\bar{\mathcal{I}}_\rmi{m}$, $\rm{m}\in\{\rm{b,d,g}\}$. 
All of these consist of 1-loop integrals of 
the type that have been thoroughly investigated in the literature, 
see e.g.~refs.~\cite{az}--\cite{jan}.
Neglecting again contact terms, 
and noting that in \eqs\nr{Gtheta_bare}, \nr{Gchi_bare}
the sum-integrals $\mathcal{I}_\rmi{b}$ and $\mathcal{I}_\rmi{g}$ are 
multiplied by terms of $\rmO(\epsilon)$, 
we obtain to the required order in $\epsilon$:
\ba
 \bar{\mathcal I}_\rmi{b}&=&\rmO(\epsilon^0)
 \;,
 \\%-\fr{\pi^2 T^7}{12} 
 %\fr{1}{\bar{x}}\fr{{\rm d}^2}{{\rm d}\bar{x}^2}
 %\bigg[\fr{1}{\bar{x}}{\rm coth}\,\bar{x}\bigg]
 %+\rmO(\epsilon) \rightarrow 
 %-\fr{\pi^2 T^7}{6}\fr{1}{\bar{x}^4}+\rmO(e^{-2\bar{x}}),\\
 \bar{\mathcal I}_\rmi{d}&=&-\fr{\pi^2 T^7}{12}
 \fr{1}{\bar{x}}\fr{{\rm d}^3}{{\rm d}\bar{x}^3}
 \!\Big({\rm coth}\, \bar{x}\Big)+\rmO(\epsilon)
 \;,\\
 % \rightarrow \rmO(e^{-2\bar{x}}),\\
 \bar{\mathcal I}_\rmi{g}&=&\fr{\pi^2 T^7}{2}
 \fr{1}{\bar{x}}\fr{{\rm d}^4}{{\rm d}\bar{x}^4}
 \biggl( \fr{ {\rm coth}\,\bar{x} }{\bar{x}} \biggr)
 \fr{1}{\epsilon}+\rmO(\epsilon^0)
 \;. \la{Ig_res}
 %\rightarrow 12\pi^2 T^7\fr{1}{\bar{x}^6}
 %\fr{1}{\epsilon}+\rmO(e^{-2\bar{x}}) .
\ea
It is worth noting that the contribution of the zero mode $q_n=r_n=0$
vanishes or is of $\rmO(\epsilon)$ in all of these cases, 
so it does not need to be separately subtracted.

%%%%%%%%%%%%%%%%%%%%%%%%%%% SUBSECTION %%%%%%%%%%%%%%%%%%%%%%%%%%%%%%
%
\subsection{$\bar{\mathcal{I}}_\rmi{f}'$}

As a warm-up for the more challenging integrals, 
let us consider $\bar{\mathcal{I}}_\rmi{f}'$ in detail. 
Defining
\ba
 \Pi(R)&\equiv&\sumint{Q}\fr{1}{Q^2(Q-R)^2} \;, \la{Pidef} 
\ea
and using a representation from 
ref.~\cite{az} (cf.\ also \eq\nr{Pi} below, with the flag
$\alpha$ set to zero and the zero-mode part 
$T\delta_{r_n}/8r$ which has already been accounted for 
in \eq\nr{If0} omitted), 
we readily obtain for the sum-integral in question
(the prime reminds of the zero-mode omission):
\ba
 \bar{\mathcal{I}}_\rmi{f}' &=& -\nabla^2\int_\vec{p} 
 e^{i\mathbf{p\cdot x}}
 \biggl\{ \sumint{R}\frac{\Pi(R)}{(p-R)^2} \biggr\}' \nn
 &=& -\nabla^2 \int_\vec{p}  
 \sumint{R}\frac{e^{i\mathbf{(p-r)\cdot x}}}{(p-R)^2}\bigg[
  \frac{e^{i\mathbf{r\cdot x}}}{(4\pi)^2}
  \left(\frac{\bar \Lambda^2}{R^2}\right)^\epsilon 
  \left( \frac{1}{\epsilon}+2 \right)\nn
 & & \hspace*{3cm}
 + \, \frac{T}{(4\pi)^2}\int\!\dd^3\vec{y}\,
     \frac{e^{i\mathbf{r\cdot (x+ y)}}}{y^2} 
     \left(\coth \bar y -\frac{1}{\bar y}-\delta_{r_n}\right) 
     e^{-|r_n|y} \bigg] \nn
 &=& -\fr{T\nabla^2}{(4\pi)^3}
     \biggl[ \bar \Lambda^{2\epsilon}\biggl( \frac{1}{\epsilon}+2 \biggr)
    \sum_{r_n}\fr{e^{-|r_n|x}}{x} \int_\vec{r} e^{i\mathbf{r\cdot x}}
    \frac{1}{\(R^2\)^{\epsilon}}
   \nn 
 & & \hspace*{3cm} 
   + \, T \sum_{r_n}\fr{e^{-2|r_n|x}}{x^3}
    \biggl(\coth \bar x -\frac{1}{\bar x}-\delta_{r_n} \biggr)\biggr] \nn
&=&-\fr{3\pi^2 T^7}{4}
    \fr{1}{\bar{x}}\fr{{\rm d}^2}{{\rm d}\bar{x}^2} 
    \biggl( \fr{1}{\bar{x}^2 \mathop{\rm sinh}^2\bar{x}}\biggr) 
    +\rmO(\epsilon)
   \;.
\ea
In the last step we expanded in $\epsilon$ and dropped contact terms; 
we also inserted the integrals 
\be
 \int_\vec{r} \frac{e^{i\mathbf{r\cdot x}}}{r_n^2 + r^2}
 = \frac{e^{-|r_n|x}}{4\pi x}
 \;, \quad
 \int_\vec{r} e^{i\mathbf{r\cdot x}} \, \ln({r_n^2 + r^2})
 = -\frac{1}{2\pi x^3} (1+|r_n|x)e^{-|r_n|x}
 \;, \la{3dF}
\ee
and carried out subsequently 
the elementary sums over $r_n$, 
\be 
 \sum_{r_n} e^{-2|r_n|x} = \mathop{\rm coth}\bar{x}
 \;, \quad 
 \sum_{r_n} |r_n| x \, e^{-2|r_n|x} = 
 \frac{\bar{x} }{ 2 \mathop{\rm sinh}^2\bar{x} } 
 \;.  \la{elesums}
\ee

%%%%%%%%%%%%%%%%%%%%%%%%%%% SUBSECTION %%%%%%%%%%%%%%%%%%%%%%%%%%%%%%%%%%
%
\subsection{$\bar{\mathcal{I}}_\rmi{h}'$}

Next, we inspect the sum-integral $\bar{\mathcal{I}}_\rmi{h}'$, 
the evaluation of which is in many ways similar to 
that of $\bar{\mathcal{I}}_\rmi{f}'$. Separating again 
the zero temperature part of $\Pi(R)$ from the rest, we obtain
\ba
 \bar{\mathcal{I}}_\rmi{h}' &=& 
 \fr{\bar{\Lambda}^{2\epsilon}}{(4\pi)^2}\left( \frac{1}{\epsilon}+2 \right) 
 \int_\vec{p} e^{i\mathbf{p\cdot x}}  
 \sumint{R}\frac{p^4}{(R^2)^{1+\epsilon}(p-R)^2}\nn
 & & \quad +\, \frac{T}{(4\pi)^2}\nabla^4\int_\vec{p} e^{i\mathbf{p\cdot x}} 
 \sumint{R}\frac{1}{R^{2}(p-R)^2}\int\!\dd^3\vec{y}\,
     \frac{e^{i\mathbf{r\cdot y}}}{y^2} 
  \left(\coth \bar y -\frac{1}{\bar y}-\delta_{r_n}\right) e^{-|r_n|y}\nn
&\equiv&  
 \bar{\mathcal{I}}_\rmi{h}^0+ \bar{\mathcal{I}}_\rmi{h}^T
 \;, \label{ih0}
\ea
where we leave the first part intact for now 
(we return to it in \se\ref{poles}) and only consider 
the second term, $\bar{\mathcal{I}}_\rmi{h}^T$. 
For this function, we obtain 
\ba
\bar{\mathcal{I}}_\rmi{h}^T&=& \frac{T^2\nabla^4 }{(4\pi)^4} 
 \biggl[\frac{1}{x} \sum_{r_n} \int\!\dd^3\vec{y}\, \frac{1}{y^2|\mathbf{x-y}|}
    \biggl(\coth \bar y -\frac{1}{\bar y}-\delta_{r_n} \biggr) 
   e^{-|r_n|(x+y+|\mathbf{x-y}|)}\biggr] \nn
&=&-\fr{\pi^2 T^7}{2 \bar{x}} \frac{{\rm d}^4}{{\rm d}\bar{x}^4} 
  \int_0^\infty \! \frac{\dd \bar{y}}{\bar{x}\bar{y}} \, \biggl\{
  \biggl(\coth \bar y -\frac{1}{\bar y}\biggr)
  \biggl[ \bar{y}_<+\ln\biggl(
    \fr{e^{2\bar{y}_>}-1}{e^{2(\bar{x}+\bar{y})}-1} \biggr)\biggr] 
  +\bar{y}_< \biggr\}
 \;, 
%  &\equiv& \pi^2T^7 f_1^\rmi{num}(\bar{x})\nn
%&\rightarrow&\pi^2 T^7\(13-12\ln(2\bar{x})\)
%\fr{1}{\bar{x}^6}+ \rmO(e^{-2\bar{x}}),
\ea
where $\bar{y}_<\equiv {\rm min}(\bar{y},\bar{x})$ 
and $\bar{y}_>\equiv {\rm max}(\bar{y},\bar{x})$. 
After taking the derivatives with respect to $\bar{x}$, 
we are left with an integral that appears to be difficult 
to perform analytically, but is numerically quite benign. 
Thus, we have opted to evaluate it numerically 
(cf.\ \se\ref{se:results}).

%%%%%%%%%%%%%%%%%%%%%%%%%%%%%%%% SUBSECTION %%%%%%%%%%%%%%%%%%%%%%%%%%%%
%
\subsection{$\bar{\mathcal{I}}_\rmi{i}'$}

As in ref.~\cite{lvv}, we reduce the evaluation of 
$\bar{\mathcal{I}}_\rmi{i}'$ to the other integrals, 
in particular $\bar{\mathcal{I}}_\rmi{i'}'$, employing the relation
\be
 \bar{\mathcal{I}}_\rmi{i}'  =
 \bar{\mathcal{I}}_\rmi{a}'+\bar{\mathcal{I}}_\rmi{e}' 
 - \bar{\mathcal{I}}_\rmi{f}'
 + \bar{\mathcal{I}}_\rmi{i'}' 
 \;. \la{Ii_red}
\ee

%%%%%%%%%%%%%%%%%%%%%%%%%%%%%%%% SUBSECTION %%%%%%%%%%%%%%%%%%%%%%%%%%%%
%
\subsection{$\bar{\mathcal{I}}_\rmi{i'}'$}

The evaluation of $\bar{\mathcal{I}}_\rmi{i'}'$ 
involves a new type of 1-loop ``self-energy'' structure, 
as instead of $\Pi(R)$ of \eq\nr{Pidef} it contains the function
\ba
 \Pi_{ij}(R)&\equiv& \sumint{Q} \frac{q_i q_j}{Q^2(Q-R)^2} 
  \;= \; A(R) \, \delta_{ij} + B(R) \, r_i r_j\, 
 \;,
\ea
where in the latter step we have used three-dimensional rotational 
symmetry to write the tensor in terms of two scalar quantities. 
As usual $A$ and $B$ can be determined by projecting the defining
equation with $\delta_{ij}$ and $r_ir_j$; moreover, by completing
squares, the results can be reduced to the known tadpole
$\Tinti{Q} \, 1/Q^2 = T^2/12$ as well as to the
self-energies (cf.\ refs.~\cite{az,jan})
\ba
  \Tint{Q} \frac{1}{Q^2(Q-R)^2} & = & 
   \frac{1}{(4\pi)^2}
   \left(\frac{\bar \Lambda^2}{R^2}\right)^\epsilon 
   \left( \frac{1}{\epsilon}+2 \right)
   + \frac{\alpha\, T^2}{6 R^2} + \frac{T\delta_{r_n}}{8 r}
   \nn
 & & \; 
  + \, \frac{T}{(4\pi)^2}\int\!\dd^3\vec{y}\,
      \frac{e^{i\mathbf{r\cdot y}}}{y^2} 
     \biggl(\coth \bar y -\frac{1}{\bar y}-\frac{\alpha\,\bar{y}}{3} 
     - \delta_{r_n}\biggr) 
     e^{-|r_n|y} 
    \;,
 \la{Pi} \\ 
  \Tint{Q} \frac{4 q_n^2 - r_n^2}{Q^2(Q-R)^2} & = & 
   - \frac{1}{(4\pi)^2}
   \left(\frac{\bar \Lambda^2}{R^2}\right)^\epsilon 
   \frac{r^2}{D-1} \left( \frac{1}{\epsilon}+2 \right) 
   + \frac{\alpha\, T^2 r_n^2}{6 R^2} 
   \nn
 & & \; 
  + \, \frac{T}{(4\pi)^2}\int\!\dd^3\vec{y}\,
      \frac{e^{i\mathbf{r\cdot y}}}{y^2} \partial_y^2 
     \biggl[ 
     \biggl(\coth \bar y -\frac{1}{\bar y}-\frac{\alpha\,\bar{y}}{3} 
      - \delta_{r_n} \biggr) 
     e^{-|r_n|y} \biggr] 
    \;. \hspace*{0.6cm}
 \la{Pi2}
\ea
The coefficient $\alpha$ serves as a ``flag'', to be chosen 
freely (below $\alpha = 1$ for $r_n\neq 0$). 
Note that in \eq\nr{Pi2} the zero-mode subtraction has actually 
no influence; $\partial_y^2 [\, 1 \,]= 0$.

Defining now, for brevity,
\ba
 f_n(y) &\equiv& \biggl(\coth \bar y -\frac{1}{\bar y} 
 -\frac{\bar y}{3} -\delta_{r_n}\biggr)e^{-|r_n|y} \;,\\
 \tilde{f}_n(y) &\equiv& \biggl(\coth \bar y -\frac{1}{\bar y} 
 -\frac{\bar y}{3} -\Bigl(1-\fr{\bar{y}}{3}\Bigr)
  \delta_{r_n}\biggr)e^{-|r_n|y}
 \;,
\ea
where in the latter case we have, for future convenience, 
removed the $\alpha$-flagged contribution in the case of 
the zero-mode, the functions $A$ and $B$ become
\ba
A(R) &=& -\frac{R^2}{4(D-1)}\frac{1}{(4\pi)^2}
   \left( \frac{\bar \Lambda^2}{R^2}\right)^\epsilon
   \left( \frac{1}{\epsilon} +2 \right) - 
   \biggl\{ \frac{T r}{64} \delta_{r_n} \biggr\} \nn
& &\; - \, \frac{R^2}{8r^2} \frac{T}{(4\pi)^2} 
  \int\!\dd^3\vec{y}\, \frac{e^{i\mathbf{r\cdot y}}}{y^2} 
       (\partial_y^2 +r^2) f_n(y) 
  \;, \label{AQ} \\
B(R) &=& \frac{D}{4(D-1)}
  \frac{1}{(4\pi)^2}\left( \frac{\bar \Lambda^2}{R^2}\right)^\epsilon
  \left( \frac{1}{\epsilon} +2 \right)   
 + \(\fr{1}{12}-\fr{\delta_{r_n}}{16}\)\frac{T^2}{R^2}
 + \biggl\{ \frac{3T}{64r} \delta_{r_n} \biggr\} \nn
 & & \; + \, 
 \frac{1}{8r^4} \frac{T}{(4\pi)^2} \int\!\dd^3\vec{y}\, 
 \frac{e^{i\mathbf{r\cdot y}}}{y^2} 
  \Bigl[ (3 R^2-2 r^2) \partial_y^2 +(R^2+2 r^2)r^2 \Bigr]
  \tilde f_n(y)
 \;.
\ea
Defining $A' \equiv A + \{ \frac{T r}{64} \delta_{r_n} \}$ 
and $B' \equiv B - \{ \frac{3T}{64r} \delta_{r_n} \}$, i.e.\ 
dropping the terms whose effects
were already accounted for in \eq\nr{Iipzero}, 
the sum-integral under study becomes
\ba
\bar{\mathcal{I}}_\rmi{i'}' &=&
 4\int_\vec{p} e^{i\mathbf{p\cdot x}} \sumint{R}
 \frac{1}{R^2(p-R)^2}\Bigl[ p^2 A'(R) +(\mathbf{p\cdot r})^2 B'(R) \Bigr] \nn
 &=& -4\nabla^2 T \sum_{r_n} \frac{e^{-|r_n|x}}{4\pi x} 
 \int_\vec{r} e^{i\mathbf{r\cdot x}} \frac{A'(R)}{R^2}
 + 4T \sum_{r_n} \int_\vec{p} \frac{e^{i\mathbf{p\cdot x}}}{p^2+r_n^2}
 \int_\vec{r} \frac{e^{i\mathbf{r\cdot x}}}{R^2} 
 (r^2+\mathbf{p\cdot r})^2 B'(R) \nn
 &=& 4T \sum_{r_n}\biggl\{ -\nabla^2 \(\frac{e^{-|r_n|x}}{4\pi x} 
 \int_\vec{r} e^{i\mathbf{r\cdot x}} \frac{A'(R)}{R^2}\)
  + \frac{\partial^2}{\partial x_i \partial x_j} 
 \biggl[ \frac{e^{-|r_n|x}}{4\pi x}
  \frac{\partial^2}{\partial x_i \partial x_j}  
 \int_\vec{r} e^{i\mathbf{r\cdot x}} \frac{B'(R)}{R^2} \biggr] \biggr\} \nn
 &\equiv& 4\(\bar{\mathcal{I}}_\rmi{i'}^A+ \bar{\mathcal{I}}_\rmi{i'}^B\)
 \;,
\ea
which defines the two functions we now set out to compute.

Let us start with $\bar{\mathcal{I}}_\rmi{i'}^A$ 
and divide it into two parts according to the two terms in \eq(\ref{AQ}),
\ba
 \bar{\mathcal{I}}_\rmi{i'}^A &\equiv& 
 \bar{\mathcal{I}}_\rmi{i'}^{A0}+\bar{\mathcal{I}}_\rmi{i'}^{AT}
 \;,
\ea
where the first ``zero-temperature'' piece reads
\ba
\bar{\mathcal{I}}_\rmi{i'}^{A0} &=&
  \fr{1}{4(D-1)}\frac{\bar{\Lambda}^{2\epsilon}}{(4\pi)^3}
  \left(\frac{1}{\epsilon} +2 \right)
    T\sum_{r_n} \nabla^2 \, \biggl[ \frac{e^{-|r_n|x}}{x} \int_\vec{r}
    \frac{e^{i\mathbf{r \cdot x}}}{(R^2)^\epsilon} \biggr] \nn
    &=&\fr{T}{6(4\pi)^4}\nabla^2 \biggl[ \fr{1}{x^4} \sum_{r_n} 
    \(1+|r_n|x\)e^{-2|r_n|x} \biggr]     \nn
    &=&-\fr{\pi^2 T^7}{48}
    \fr{1}{\bar{x}}\fr{{\rm d}^3}{{\rm d}\bar{x}^3} 
    \biggl( \fr{{\rm coth}\,\bar{x}}{\bar{x}^2}
    \biggr) 
    \;.
%\rightarrow \fr{\pi^2 T^7}{2}\fr{1}{\bar{x}^6}+\rmO(e^{-2\bar{x}}) .
\ea
In the second step we expanded in $\epsilon$, 
dropped contact terms, and made use of \eq\nr{3dF}; in the last
step the sums were carried out like in \eq\nr{elesums}.
The finite temperature part can, on the other hand, 
be written as
\ba
 \bar{\mathcal{I}}_\rmi{i'}^{AT} &=& 
 \frac{T^2}{8(4\pi)^3} \nabla^2 \sum_{r_n} \frac{e^{-|r_n|x}}{x}
 \int\! \frac{\dd^3\vec{y}}{y^2} \,
 \int_\vec{r} e^{i\mathbf{r}\cdot (\mathbf{x}+\mathbf{y})}
    \left( 1 +\frac{1}{r^2}\partial_y^2 \right) f_n(y) \nn
&=& \frac{T^2}{8(4\pi)^3} \nabla^2 
  \sum_{r_n} \frac{e^{-|r_n|x}}{x} \left[
    \frac{f_n(x)}{x^2} + 
    \int\!\dd^3\vec{y}\, \frac{1}{4\pi y^2|\mathbf{x+y}|} 
    \partial_y^2 f_n(y) \right] \nn
&=& \frac{T^2}{8(4\pi)^3} \nabla^2 \sum_{r_n} 
    \frac{e^{-|r_n|x}}{x} \left[
    \frac{f_n(x)}{x^2} + 
    \int_0^\infty \!\dd y\, \frac{1}{y_>} 
    \partial_y^2 f_n(y) \right] \nn
&=& \frac{T^2}{4(4\pi)^3} \nabla^2 \sum_{r_n} 
    \frac{e^{-|r_n|x}}{x} 
    \int_x^\infty \! \dd y\, \frac{f_n(y)}{y^3}  \nn
&=& \frac{\pi^2 T^7}{8 \bar{x}}\fr{{\rm d}^2}{{\rm d}\bar{x}^2} 
    \int_{\bar{x}}^\infty \! 
    \frac{\dd \bar{y}}{\bar{y}^3}
    \, \biggl[ \coth\biggl(\frac{\bar x+\bar y}{2}\biggr)
    \biggl(\coth \bar y -\frac{1}{\bar y} -\frac{\bar y}{3}
    \biggr) -1 \biggr]
  \;,
%&\equiv& \pi^2T^7 f_2^\rmi{num}(\bar{x})\nn
%&\rightarrow& -\fr{\pi^2 T^7}{12}
% \(\fr{1}{\bar{x}^4}+\fr{6}{\bar{x}^6}\)+\rmO(e^{-2\bar{x}}) ,
\ea
where the last integral again seems to require numerical evaluation. 

Moving then on to $\bar{\mathcal{I}}_\rmi{i'}^B$, 
we once more split the integral into two parts, 
now according to the $T=0$ and $T\neq 0$ terms of $B'(Q)$. This produces
\ba
 \bar{\mathcal{I}}_\rmi{i'}^B &\equiv& 
 \bar{\mathcal{I}}_\rmi{i'}^{B0}+\bar{\mathcal{I}}_\rmi{i'}^{BT}, \nn
 \bar{\mathcal{I}}_\rmi{i'}^{B0} &=&
 \fr{D}{4(D-1)}\frac{\bar{\Lambda}^{2\epsilon}}{(4\pi)^2}
 \left(\frac{1}{\epsilon} +2 \right)\int_\vec{p} e^{i\mathbf{p\cdot x}}\, 
  \sumint{R}
    \frac{(\mathbf{p\cdot r})^2}{(R^2)^{1+\epsilon} (p-R)^2}
 \;, \label{iipb0}
\ea
which we analyse together with the other divergent terms
in \se\ref{poles}. 
With $\bar{\mathcal{I}}_\rmi{i'}^{BT}$, on the other hand,
we encounter a similar albeit more complicated expression as 
with $\bar{\mathcal{I}}_\rmi{i'}^{AT}$. Its evaluation 
nevertheless proceeds along the same lines: One begins 
by performing the spatial $\mathbf{r}$ integral, then takes 
care of the angular part of the $\mathbf{y}$ integration,
partially integrates, and finally performs the sum over 
$r_n$ (which we leave intact for the moment). 
The result of the procedure reads 
(a prime indicating the omission of the zero-mode)
\ba
\bar{\mathcal{I}}_\rmi{i'}^{BT} &=& 
  \frac{T}{4\pi} 
  {\sum_{r_n}}' \frac{\partial^2}{\partial x_i \partial x_j} 
  \frac{e^{-|r_n|x}}{ x}
    \frac{\partial^2}{\partial x_i \partial x_j} 
    \biggl\{
    \frac{T^2}{12} \frac{e^{-|r_n|x}}{8\pi|r_n|}\nn
& & \; -\, 
    \frac{T}{8(4\pi)^2}\frac{2}{|r_n^3| x} \int_0^\infty\!\frac{\dd y}{y^3}
        \left( 1 +|r_n|y +r_n^2 y^2\right)e^{-|r_n|(x+y)}  f_n(y)  \nn
& & \; + \,
    \frac{T}{8(4\pi)^2}\frac{2}{|r_n^3| x} \int_0^x \!\frac{\dd y}{y^3}
        \left( 1 -|r_n|y +r_n^2 y^2\right)e^{-|r_n|(x-y)} f_n(y) \nn
& & \; + \, 
   \frac{T}{8(4\pi)^2}\int_x^\infty\!\frac{\dd y}{y^3} \left[
      y^2-x^2-\frac{4}{r_n^2} 
   +\frac{2}{|r_n^3| x}\left( 1 +|r_n|y +r_n^2 y^2\right)
   e^{-|r_n|(y-x)} \right] f_n(y) 
  \biggr\}\nn
&+&\frac{T^2}{8(4\pi)^3} \frac{\partial^2}{\partial x_i \partial x_j} 
   \frac{1}{x}
    \frac{\partial^2}{\partial x_i \partial x_j} 
    \biggl\{
    \frac{8}{3x}\int_0^x \!\dd y\,\tilde f_0(y)
    +\int_x^\infty \!\dd y\,
    \left(\frac{3}{y}-\frac{x^2}{3y^3}\right)\tilde f_0(y) 
    \biggr\}
 \;. 
%&\equiv& \pi^2T^7 f_3^\rmi{num}(\bar{x})\nn
%&\rightarrow& -\pi^2 T^7\,\fr{\ln(2\bar{x})-9/4}{\bar{x}^6}.\nonumber
\ea
The derivatives can be taken by noting that the independence of 
the above expressions on angular variables implies
\ba
 \partial_i\partial_j \Bigl[ g(x)\partial_i\partial_j f(x) \Bigr]
 &=& \frac{1}{x^2} \partial_x^2 \left(x^2 g f'' \right)
    -\frac{2}{x^2} \partial_x\( g f' \)
 \;. \label{fgder}
\ea
Then the first row can be simplified into 
\ba
 & & \hspace*{-1.5cm}
  \frac{T}{4\pi} 
  {\sum_{r_n}}' \frac{\partial^2}{\partial x_i \partial x_j} 
  \biggl\{ \frac{e^{-|r_n|x}}{ x}
    \frac{\partial^2}{\partial x_i \partial x_j} 
    \biggl[ \,  
    \frac{T^2}{12} \frac{e^{-|r_n|x}}{8\pi|r_n|}
    \, \biggr] \biggr\} \nn 
 & = & 
 \frac{\pi^2 T^7}{3}
 \biggl[ 
   -\frac{n(2\bar{x})}{2\bar{x}^4}
   +\frac{n'(2\bar{x})}{\bar{x}^3}
   -\frac{n''(2\bar{x})}{\bar{x}^2}
   -\frac{n^{(3)}(2\bar{x})}{\bar{x}}
 \biggr] 
 \;, 
\ea
where $n$ stands for the bosonic distribution 
function $n(x) \equiv 1/(e^x-1)$,  
whereas the last row (the contribution from $r_n=0$, $q_n\neq 0$) becomes
\ba
 & & \hspace*{-1.5cm}
   \frac{T^2}{8(4\pi)^2} \frac{\partial^2}{\partial x_i \partial x_j} 
   \biggl\{ \frac{1}{4\pi x}
    \frac{\partial^2}{\partial x_i \partial x_j} 
    \biggl[ 
    \frac{8}{3x}\int_0^x \!\dd y\,\tilde f_0(y)
    +\int_x^\infty \!\dd y\,
    \left(\frac{3}{y}-\frac{x^2}{3y^3}\right)\tilde f_0(y) 
    \biggr] \biggr\} \nn
 & = & \frac{\pi^2T^7}{4}
   \biggl[
     \frac{9}{\bar{x}^6} 
   + \frac{4}{\bar{x}^6} 
     \ln\biggl( \frac{1-e^{-2\bar{x}}}{2\bar{x}} \biggr) 
    - \frac{11\, n(2\bar{x})}{\bar{x}^5}
    + \frac{10\, n'(2\bar{x})}{\bar{x}^4}
    - \frac{4\, n''(2\bar{x})}{\bar{x}^3}
   \biggr]
 \;.
\ea
The other parts contain integrals that we collect 
together in \se\ref{se:results} for numerical evaluation. 

%%%%%%%%%%%%%%%%%%%%%%%%%%%%% SUBSECTION %%%%%%%%%%%%%%%%%%%%%%%%%%%%%%%%%%%
%
\subsection{$\bar{\mathcal{I}}_\rmi{j}'$}
\label{se:Ij}

From the numerics point of view, 
the sum-integral $\bar{\mathcal{I}}_\rmi{j}'$ appears to be 
the most non-trivial one, as there are not many analytic 
tricks that we can apply to it. The most important observation 
is that $\bar{\mathcal{I}}_\rmi{j}'$ 
is both UV and IR finite, implying that we can 
set $\epsilon=0$ and consider its evaluation in three-dimensional 
coordinate space. Fourier transforming all 
the propagators and subsequently rescaling 
the radial coordinates by $x$, we obtain
\ba
 \bar{\mathcal{I}}_\rmi{j}'
  &=& -\frac{T^2\nabla^6}{(4\pi)^5}  \int \!\dd^3\vec{y}\, \dd^3\vec{z}\,
  \frac{
   {\sum_{q_n,r_n}'}
    e^{ - |q_n|(y+z)- |r_n|(|\mathbf{x-y}|+|\mathbf{x-z}|)
   -|q_n-r_n||\mathbf{y-z}| }
  }{yz|\mathbf{x-y}||\mathbf{x-z}||\mathbf{y-z}|}
  \nn
  &=& -\frac{\pi T^7}{16 \bar{x}}
    \int_0^\infty \!\! \dd \tilde{y}\, \dd \tilde{z} 
    \int_{-1}^1 \!\! \dd t_1\: \dd t_2 \int_0^{2\pi} \!\!\dd\phi\;
     \frac{\tilde{y}\tilde{z}}
     {|\mathbf{e-\tilde{y}}|
      |\mathbf{e-\tilde{z}}|
      |\mathbf{\tilde{y}-\tilde{z}}|} \nn
     &&\hspace*{2cm} \times
  \frac{\dd^6}{\dd \bar{x}^6}  
  \biggl\{ \bar{x}^2{\sum_{m,n}}'
  e^{-\bar{x} \bigl[ |m|(\tilde{y}+\tilde{z})+
  |n|(|\mathbf{e-\tilde{y}}|+|\mathbf{e-\tilde{z}}|)+
  |m-n||\mathbf{\tilde{y}-\tilde{z}}| \bigr] }
  \biggr\} 
 \;,
\ea
where $\vec{e}$ is a unit vector and 
the prime on the sum means that the simultaneous 
zero mode $m=n=0$ has been left out. This sum can now be 
explicitly carried out, giving
\ba
 \bar{\mathcal{I}}_\rmi{j}' &=&
 -\frac{\pi T^7}{8 \bar{x}}
    \int_0^\infty \!\! \dd \tilde{y}\: \dd \tilde{z} 
  \int_{-1}^1 \!\! \dd t_1\: \dd t_2 \int_0^{2\pi} \!\!\dd\phi\;
     \frac{\tilde{y}\tilde{z}}
         {|\mathbf{e-\tilde{y}}|
          |\mathbf{e-\tilde{z}}|
          |\mathbf{\tilde{y}-\tilde{z}}|} \\
 & & \; \times \, \frac{\dd^6}{\dd \bar{x}^6}  \Bigg\{\bar{x}^2
    \Big[ n(\bar xa) +n(\bar xb)+n(\bar xc)
     +n(\bar xa)n(\bar xb)+n(\bar xb)n(\bar xc)
     +n(\bar xc)n(\bar xa)\Big] \Bigg\}
 \;, \nonumber
%&\equiv& \pi^2T^7 f_4^\rmi{num}(\bar{x}),\nonumber
\ea
where we have defined 
the combinations 
$
 a \equiv \tilde{y}+\tilde{z}+|\mathbf{e-\tilde{y}}|
 +|\mathbf{e-\tilde{z}}|
$, 
$
 b \equiv \tilde{y}+\tilde{z}+|\mathbf{\tilde{y}-\tilde{z}}|
$ 
and 
$
 c \equiv |\mathbf{e-\tilde{y}}|
 +|\mathbf{e-\tilde{z}}|+|\mathbf{\tilde{y}-\tilde{z}}|
$. 
The sixth derivative of the expression on the last line contains 
a large number of terms, but its evaluation is straightforward 
to automatize using the relation $n'(x)=-n(x)[1+n(x)]$. 
The remaining 5d integral can be performed 
using numerical methods.

%
%In the limit of large $x$, the integral obtains its dominant
% contributions, when one of the Matsubara frequencies vanishes,
% i.e.~$q_0=0$ or $r_0=0$, in which case there is no exponential 
%suppression. We obtain then the integral
%\ba
%\bar{\mathcal{I}}_\rmi{j}^\rmi{IR} &=&
% -4\frac{T^2}{(4\pi)^5}\nabla^6  \int \!\dd^3\vec{y}\, \dd^3\vec{z}\,
%    \frac{1}{yz|\mathbf{x-y}||\mathbf{x-z}||\mathbf{y-z}|}
%\fr{1}{e^{\bar{y}+\bar{z}+|\mathbf{\bar{y}-\bar{z}}|}-1},
%\ea
%which seems to be quite difficult to perform analytically. 
%If we, however, expand the factors $|\mathbf{x-y}|$ 
%and $|\mathbf{x-z}|$ in the denominator in the limit 
%of large $x$ in order to obtain the leading order terms 
%in a $1/x$ expansion, we readily find the result
%\ba
%f_4^\rmi{num}(\bar{x})&\rightarrow& 
%-\fr{180\zeta(3)}{\bar{x}^8}-\fr{1260\zeta(5)}
%{\bar{x}^{10}} +\rmO(1/\bar{x}^{12}).
%\ea
%

%%%%%%%%%%%%%%%%%%%%%%%%% SUBSECTION %%%%%%%%%%%%%%%%%%%%%%%%%%
%
\subsection{Divergent terms}
\label{poles}

Above, we have left two integrals unevaluated, 
namely the UV-divergent term $\bar{\mathcal{I}}_\rmi{h}^0$ 
of \eq(\ref{ih0}) and $\bar{\mathcal{I}}_\rmi{i'}^{B0}$ 
of \eq(\ref{iipb0}). These contributions are expected 
to partially cancel against the renormalization 
of the gauge coupling in connection with 
the 1-loop terms in \eqs(\ref{Gtheta_bare}) 
and (\ref{Gchi_bare}). Here, we evaluate the sum 
of these three functions, obtaining a UV-finite result.

Inspecting the forms of $\bar{G}_\theta(x)$ and $\bar{G}_\chi(x)$ 
in \eqs(\ref{Gtheta_bare}) and (\ref{Gchi_bare}), we observe that 
the only parts containing divergences are 
\be
   \frac{\bar{G}_\theta^\rmi{div}}{4 d_A c_\theta^2}
 = \frac{\bar{G}_\chi^\rmi{div}}{-16 d_A c_\chi^2 (D-3)}
 = (D-2) 
 \biggl[
  \frac{\gB^4}{2} \bar{\mathcal{J}}_\rmi{b} 
 + \gB^6 \Nc 
   \Bigl( 
     3 \bar{\mathcal{I}}_\rmi{h}^0 + 
     2\times 4 
    \bar{\mathcal{I}}_\rmi{i'}^{B0} \Bigr)
 \biggr] 
 \;.
\ee
Expanding $\gB^2$ according to \eq\nr{gB}, we are left 
to consider the combination
\ba
 \bar{\mathcal{I}}_\rmi{div} & \equiv &
 -\fr{1}{(4\pi)^2}\fr{11}{3\epsilon}\bar{\mathcal{J}}_b
 +3\bar{\mathcal{I}}_\rmi{h}^0
 +8\bar{\mathcal{I}}_\rmi{i'}^{B0}\nn
 &=& \fr{1}{(4\pi)^2} 
  \int_\vec{p} e^{i\mathbf{p\cdot x}} 
  \sumint{R}\frac{1}{R^2(p-R)^2}
  \bigg\{-\fr{11p^4}{3\epsilon}\nn
 && \hspace*{1.5cm}
  +\, \left(\frac{1}{\epsilon}+2 \right)
  \(\fr{\bar{\Lambda}^2}{R^2}\)^{\epsilon}
  \(3p^4+\fr{2D}{D-1}(\mathbf{p\cdot r})^2\)\bigg\}\nn
&=&\fr{2}{3(4\pi)^2}
 \int_\vec{p} e^{i\mathbf{p\cdot x}} 
 \sumint{R}\frac{1}{R^2(p-R)^2}
 \bigg\{\fr{4(\mathbf{p\cdot r})^2-p^4}{\epsilon}\label{gthetachi} \nn
&& \hspace*{1.5cm}
 +\, 9p^4+\fr{26(\mathbf{p\cdot r})^2}{3}
 +\(\fr{9p^4}{2}+4(\mathbf{p\cdot r})^2\)
 \ln\fr{\bar{\Lambda}^2}{R^2}\bigg\}
 \;.
\ea
Writing the scalar products 
$(\mathbf{p\cdot r})^2$ in terms of $(p-R)^2$, 
we can massage the result into a relatively simple form; 
in particular, apart from contact terms, the $1/\epsilon$
divergence disappears in the course of the $\vec{r}$-integration.  
Upon denoting
\ba
 D_a(x,r_n)&\equiv&
 \int\! \fr{{\rm d}^{3}\vec{r}}{(2\pi)^{3}} 
 e^{i\mathbf{r\cdot x}} \fr{1}{(R^2)^a}
 \;,\\
 \widetilde{D}_a(x,r_n)&\equiv&
 \int\! \fr{{\rm d}^{3}\vec{r}}{(2\pi)^{3}} 
 e^{i\mathbf{r\cdot x}}\, \fr{\ln(\bar{\Lambda}^2/R^2)}{(R^2)^a}
 \;,
\ea
the remaining finite result becomes
\ba
 \bar{\mathcal{I}}_\rmi{div}
  &=&
 \fr{2T}{3(4\pi)^2}\sum_{r_n}
 \biggl\{\nabla^4 \biggl[D_1(x,r_n)\bigg( \fr{67}{6}D_1(x,r_n)
 +\fr{11}{2}\widetilde{D}_1(x,r_n) \biggr)\biggr]\nn
 &&\quad\quad\quad\quad\quad
 -\, 2\nabla^2 \Bigl[ D_1(x,r_n)
 \widetilde{D}_0(x,r_n) \Bigr]
 +D_1(x,r_n) \widetilde{D}_{-1}(x,r_n) \biggr\}
 \;.
\ea
The various functions that appear here have the forms
\ba
 D_1(x,r_n)&=&\fr{e^{-|r_n|x}}{4\pi x}
 \;,\\
 \widetilde{D}_{1}(x,r_n)&=&\fr{e^{-|r_n|x}}{4\pi x}
 \bigg[
   2\ln\(e^{\gammaE}\bar{\Lambda}x \)+
   |r_n| x \int_0^\infty \! {\rm d}\tilde{y} \, e^{-\tilde{y} |r_n|{x}} 
    \ln\biggl( \frac{\tilde{y}}{2+\tilde{y}} \biggr)
   % e^{2|r_n|x}{\rm Ei}(-2|r_n|x)
 \bigg]
 \;,\\
 \widetilde{D}_0(x,r_n)&=&\fr{(1+|r_n|x)e^{-|r_n|x}}{2\pi x^3}
 \;,\\
 \widetilde{D}_{-1}(x,r_n)&=&
 - \fr{(3+3|r_n|x+r_n^2x^2)e^{-|r_n|x}}{\pi x^5}
 \;,
\ea
which at the end leads to the result
\ba
 \bar{\mathcal{I}}_\rmi{div}
 &=&\pi^2 T^7
 \Biggl\{\fr{44\ln\(e^{\gammaE}\bar{\Lambda}x\)-57}
 {\bar{x}^6}
 -\fr{1}{3\bar{x}}\fr{{\rm d}^2}{{\rm d} \bar{x}^2}
 \biggl[ \fr{ n(2\bar{x}) }{\bar{x}^3} \biggr]
 +\fr{2}{3\bar{x}}\fr{{\rm d}^3}{{\rm d} \bar{x}^3}
 \biggl[ \fr{ n(2\bar{x}) }{\bar{x}^2} \biggr] \nn 
 & &\; + \, 
 \fr{1}{\bar{x}}\fr{{\rm d}^4}{{\rm d} \bar{x}^4}
 \bigg[\fr{67+66\ln\(e^{\gammaE}\bar{\Lambda}x\)}{18\bar{x} (e^{2\bar{x}}-1) } 
 +\fr{33}{18}
 \int_{0}^\infty \! {\rm d}\tilde{y} \, \ln\biggl( 
 \frac{2+\tilde{y}}{\tilde{y}} \biggr)
 n'\Bigl((2 + \tilde{y})\bar{x}\Bigr)
 \bigg]\Biggr\} 
 \;. \hspace*{1cm}
\ea

\vspace*{5cm}

\pagebreak

%%%%%%%%%%%%%%%%%%%%%%%%%%% SECTION %%%%%%%%%%%%%%%%%%%%%%%%%
%
\section{Basic results}
\la{se:results}

%%%%%%%%%%%%%%%%%%%%%%%%% SUBSECTION %%%%%%%%%%%%%%%%%%%%%%%%%
%
\subsection{General expressions}

We are now ready to write down our final results for 
the full theory contributions to the correlators 
$\bar G^\rmi{F}_\theta(x)$ and $\bar G^\rmi{F}_\chi(x)$, 
from \eqs(\ref{Gtheta_bare}), (\ref{Gchi_bare}) and 
\nr{Ii_red}. 
Omitting contact terms and collecting terms up to 
$\rmO(\epsilon^0)$, we obtain
\ba
  \frac{\bar{G}^\rmi{F}_\theta}{4 d_A c_\theta^2} & = & 
  g^4 \bar{\mathcal{J}}_\rmi{b} 
  + 2 g^6 \Nc \Bigl[ 
    \bar{\mathcal{I}}_\rmi{div}
    - 2 \bar{\mathcal{I}}_\rmi{d}
    - 3 \bar{\mathcal{I}}_\rmi{f}
    + 3 \bar{\mathcal{I}}_\rmi{h}
    + 2 \bar{\mathcal{I}}_\rmi{i'}
    - \bar{\mathcal{I}}_\rmi{j}
  \Bigr] 
  \;, \la{G_theta_res} \\ 
  \frac{\bar{G}^\rmi{F}_\chi}{-16 d_A c_\chi^2} & = & 
  g^4 \bar{\mathcal{J}}_\rmi{b} 
  + 2 g^6 \Nc \Bigl[ 
    \bar{\mathcal{I}}_\rmi{div}
    - 2 \bar{\mathcal{I}}_\rmi{d}
    - 3 \bar{\mathcal{I}}_\rmi{f}
    + 3 \bar{\mathcal{I}}_\rmi{h}
    + 2 \bar{\mathcal{I}}_\rmi{i'}
    - \bar{\mathcal{I}}_\rmi{j} + 2 \epsilon \bar{\mathcal{I}}_\rmi{g} 
  \Bigr] 
  \;. \la{G_chi_res}
\ea
It is immediately clear that the two channels only differ through 
a single term, the last one in \eq\nr{G_chi_res}. Inserting the 
result from \eq\nr{Ig_res}, the difference has a simple form; 
it is shown explicitly in \eqs\nr{reschi}, \nr{fLO} below 
and is displayed numerically in \fig\ref{fig:basic_a}.

Adding up the various parts from \se\ref{se:naive}
and defining the functions
\ba
 n(x) = \fr{1}{e^x-1}\;=\;\sum_{n=1}^\infty e^{-nx}
 \;,
  & & 
 i(x) \;\equiv\; \ln(1-e^{-x})\;=\;-\sum_{n=1}^\infty \fr{e^{-nx}}{n}
 \;,\\
 {\rm Li}_2(e^{-x}) \;\equiv\; \sum_{n=1}^\infty \fr{e^{-nx}}{n^2}
 \;,
  & & 
 {\rm Li}_3(e^{-x}) \;\equiv\; \sum_{n=1}^\infty \fr{e^{-nx}}{n^3}
 \;,
\ea
we end up with the expressions
\ba
 \frac{\bar{G}^\rmi{F}_\theta(x)}{4 d_A c_\theta^2}&=&
  g^4 \, 4\pi^4 T^7 \, \phi_\rmi{LO} (\bar{x}) \nn
 &+& g^6\Nc \, \pi^2 T^7
 \; \biggl\{
 \frac{11}{3} \, \phi_\rmi{LO} (\bar{x}) \,
 \ln\(\fr{e^{\gammaE} \bar{\Lambda}\bar{x}}{2\pi T}\)
 + \phi_\rmi{NLO}(\bar{x}) + \phi_1(\bar{x}) + \phi_2(\bar{x})\biggr\}
 \;, \label{restheta}  \hspace*{0.6cm} \\[3mm]
%%%%%%%
 \frac{\bar{G}^\rmi{F}_\chi(x)}
 {-16 d_A c_\chi^2} 
 &=&\frac{\bar{G}^\rmi{F}_\theta(x)}{4 d_A c_\theta^2}
 +g^6\Nc \, 2 \pi^2 T^7 \, \phi_\rmi{LO} (\bar{x})
 \;. \label{reschi}
\ea
The function $\phi_\rmi{LO}$, which determines the leading-order 
behaviour, the scale dependence, as well as the difference between
the two channels, is given by (cf.\ \eq\nr{resJb})
\ba
 \phi_\rmi{LO}(\bar{x}) & = & 
 \fr{24}{\bar{x}^6}
  +\fr{2}{\bar{x}}\fr{{\rm d}^4}{{\rm d}\bar{x}^4}
   \(\fr{n(2\bar{x})}{\bar{x}}\)
 \la{fLO} \\ 
 & = & 
 \left\{
   \begin{array}{lll} 
     \displaystyle
     \frac{120}{\bar{x}^7} + \rmO\Bigl(\frac{1}{\bar{x}}\Bigr)
    & , & 
    \bar{x} \ll 1 \\[3mm] 
     \displaystyle
     \frac{24}{\bar{x}^6} + \rmO\Bigl(e^{-2\bar{x}} \Bigr) 
    & , & 
    \bar{x} \gg 1
   \end{array} 
 \right.
 \;. \la{fLO_asy}
\ea
The function $\phi_\rmi{NLO}(\bar{x})$, which determines the 
long-distance asymptotics of the next-to-leading order correction, 
reads
\ba
 \phi_\rmi{NLO}(\bar{x}) & = & 
 -\fr{4}{3\bar{x}^4}
 +\fr{24}{\bar{x}^5}
 +\fr{88}{\bar{x}^6}\Big[i(2\bar{x})-\ln(2\bar{x})\Big]
 \la{fNLO}
 \\ 
 & = & 
 \left\{
   \begin{array}{lll} 
     \displaystyle
     -\frac{64}{\bar{x}^5} + 
     \frac{40}{3\bar{x}^4} + \rmO\Bigl(\frac{1}{\bar{x}^2}\Bigr)
    & , & 
    \bar{x} \ll 1 \\[3mm] 
     \displaystyle
     -\frac{4}{3\bar{x}^4} + \frac{24}{\bar{x}^5} - 
     \frac{88 \ln (2\bar{x})}{\bar{x}^6} 
    + \rmO\Bigl(e^{-2\bar{x}} \Bigr) 
    & , & 
    \bar{x} \gg 1
   \end{array} 
 \right.
 \;. \la{fNLO_asy}
\ea
The remaining functions $\phi_1$ and $\phi_2$ are subdominant
at long distances ($\phi_1$ by exponential terms, $\phi_2$
by powerlike terms), and take the respective forms
\ba
 \phi_1(\bar{x})&=&
  \fr{80}{3\bar{x}^4}n(2\bar{x})
 +\bigg[\fr{268}{\bar{x}^5}-\fr{24}{\bar{x}^3}\bigg]n'(2\bar{x})
 +\bigg[-\fr{676}{3\bar{x}^4}+\fr{80}{3\bar{x}^2}\bigg]n''(2\bar{x})\nn
 &+&\bigg[\fr{496}{9\bar{x}^3}-\fr{16}{\bar{x}}\bigg]n^{(3)}(2\bar{x})
 +\fr{1072}{9\bar{x}^2}n^{(4)}(2\bar{x})\nn
 &+&\fr{33}{9\bar{x}}\int_0^\infty{\rm d}\tilde{y}\, (2+\tilde{y})^4
 \ln\(\fr{2+\tilde{y}}{\tilde{y}}\)n^{(5)}
 \Bigl((2+\tilde{y})\bar{x}\Bigr)
 \nn
%\nonumber\ea\ba
&+&\fr{1}{\bar{x}}\int_0^{\bar{x}}{\rm d}\bar{y}
 \(\coth \bar y -\frac{1}{\bar y} -\frac{\bar{y}}{3}\)
 \fr{{\rm d}^4}{{\rm d}\bar{x}^4}
 \Biggl[
     \fr{{\rm Li}_3\(e^{-2\bar{x}}\)
          -{\rm Li}_3\(e^{-2(\bar{x}+\bar{y})}\)}{\bar{x}\bar{y}^3}\nn
& & \; -\,\fr{{\rm Li}_2\(e^{-2\bar{x}}\)
      +{\rm Li}_2\(e^{-2(\bar{x}+\bar{y})}\)}{\bar{x}\bar{y}^2}
  +4\,\fr{i(2(\bar{x}+\bar{y}))
         -i(2\bar{x})}{\bar{x}\bar{y}}\Biggr]\nn
&+& \fr{1}{\bar{x}}\int_{\bar{x}}^\infty{\rm d}\bar{y}
  \(\coth \bar y -\frac{1}{\bar y} -\frac{\bar{y}}{3}\)
  \fr{{\rm d}^4}{{\rm d}\bar{x}^4}
 \Biggl[ \fr{{\rm Li}_3\(e^{-2\bar{y}}\)
                -{\rm Li}_3\(e^{-2(\bar{x}+\bar{y})}\)}{\bar{x} \bar{y}^3}\nn
& & \; +\, \fr{{\rm Li}_2\(e^{-2\bar{y}}\)
      -{\rm Li}_2\(e^{-2(\bar{x}+\bar{y})}\)}{\bar{x}\bar{y}^2}
    +4\,\fr{i(2(\bar{x}+\bar{y}))
           -i(2\bar{y})}{\bar{x}\bar{y}}\Biggr]
 \;,  \la{f1_def} \\[3mm]
 \phi_2(\bar{x})&=&\frac{1}{4\pi\bar{x}}
    \int_0^\infty \!\! \dd \tilde{y}\: 
    \dd \tilde{z} \int_{-1}^1 \!\!  
    \dd t_1\: \dd t_2 \int_0^{2\pi} \!\!\dd\phi\;
    \frac{\tilde{y}\tilde{z}}
   {|\mathbf{e-\tilde{y}}|
    |\mathbf{e-\tilde{z}}|
    |\mathbf{\tilde{y}-\tilde{z}}|} \nn
 & & \; \times \, 
   \frac{\dd^6}{\dd \bar{x}^6}  \biggl\{ \bar{x}^2
    \Bigl[n(\bar{x} a) % n(\bar\alpha+\bar\beta)
   +2 n(\bar{x} b) %n(\bar\beta+\bar\gamma)
   +n(\bar{x}c)n(\bar{x}b) %n(\bar\alpha+\bar\gamma)n(\bar\beta+\bar\gamma)
   +2 n(\bar{x}a)n(\bar{x}b) %n(\bar\alpha+\bar\beta)n(\bar\beta+\bar\gamma)
   \Bigr] \biggr\}
 \;, \la{f2_def}\\
 a \!\! &\equiv& \!\! \tilde{y}+\tilde{z}+|\mathbf{e-\tilde{y}}|
 +|\mathbf{e-\tilde{z}}|
 \;, \quad
 b \equiv \tilde{y}+\tilde{z}+|\mathbf{\tilde{y}-\tilde{z}}|
 \;, \quad
 c \equiv a+b-2(\tilde{y}+\tilde{z})
% |\mathbf{e-\tilde{y}}|
% +|\mathbf{e-\tilde{z}}|+|\mathbf{\tilde{y}-\tilde{z}}|
% \\
% \bar\alpha&\equiv& 
% \bar{x}\(|\mathbf{e-\tilde{y}}|+|\mathbf{e-\tilde{z}}|\)
% \;,\quad 
%  \bar\beta \;\equiv\; \bar{x}\(\hat{y}+\hat{z}\)
% \;,\quad 
% \bar\gamma\;\equiv\; \bar{x}|\mathbf{\tilde{y}-\tilde{z}}|
 \;. \nonumber
\ea
The integrals still remaining in the definitions 
of $\phi_1$ and $\phi_2$ are carried out numerically.
The analytically known functions $\phi_\rmi{LO}$, $\phi_\rmi{NLO}$
are illustrated in \fig\ref{fig:basic_a}, the numerically
determined functions $\phi_1$, $\phi_2$ 
in \fig\ref{fig:basic_b}. 

%%%%%%%%%%%%%%%%%%%%%%%%%%%%%%%%% FIGURE %%%%%%%%%%%%%%%%%%%%%%%%%%%%%%%%%
\begin{figure}[t]

%\vspace*{-3cm}

\centerline{%
 \epsfysize=7.5cm\epsfbox{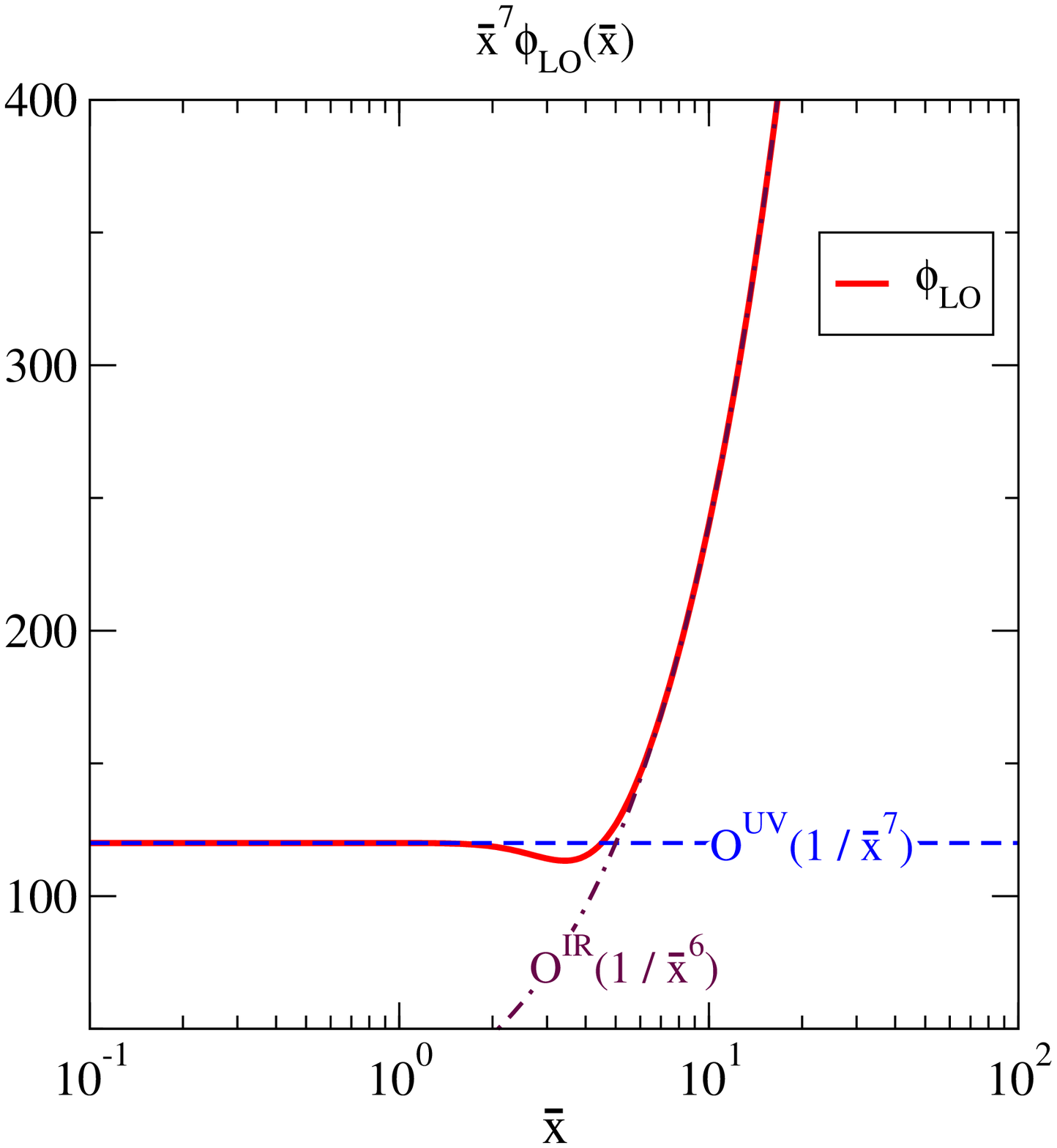}%
~~~\epsfysize=7.5cm\epsfbox{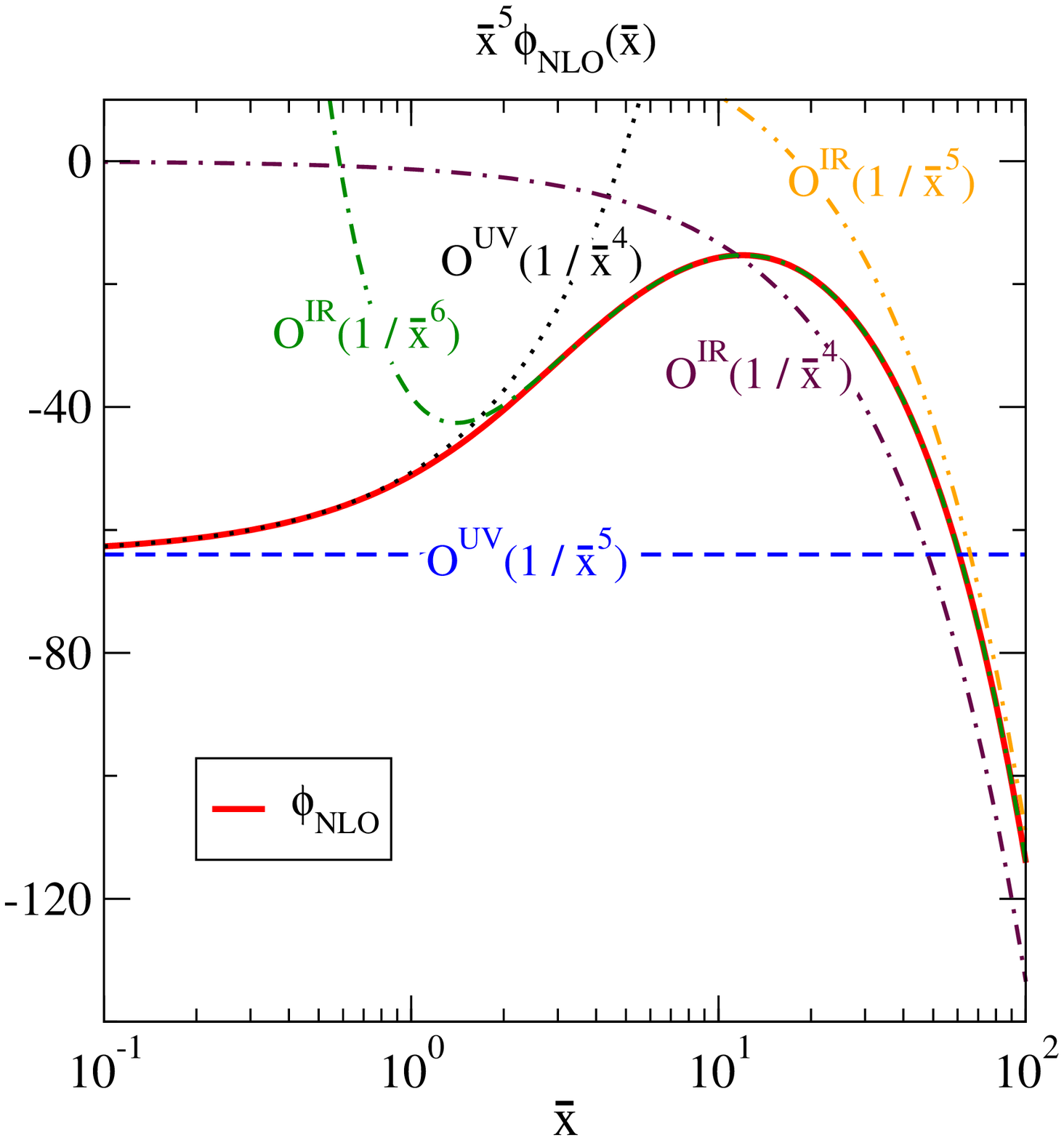}
}

\caption[a]{\small
The basic functions defined in \eqs\nr{fLO}, \nr{fNLO}, 
together with their asymptotic
limits as given in \eqs\nr{fLO_asy},
\nr{fNLO_asy}.
The functions $O^\rmii{UV/IR}$ indicate 
the nature of the limit, and the argument shows the 
order up to which terms have been included. 
}
\la{fig:basic_a}
\end{figure}
%%%%%%%%%%%%%%%%%%%%%%%%%%%%%%%%%%%%%%%%%%%%%%%%%%%%%%%%%%%%%%%%%%%%%%%%%%%

%%%%%%%%%%%%%%%%%%%%%%%%%%%%%%%%% FIGURE %%%%%%%%%%%%%%%%%%%%%%%%%%%%%%%%%
\begin{figure}[t]

%\vspace*{-3cm}

\centerline{%
 \epsfysize=7.5cm\epsfbox{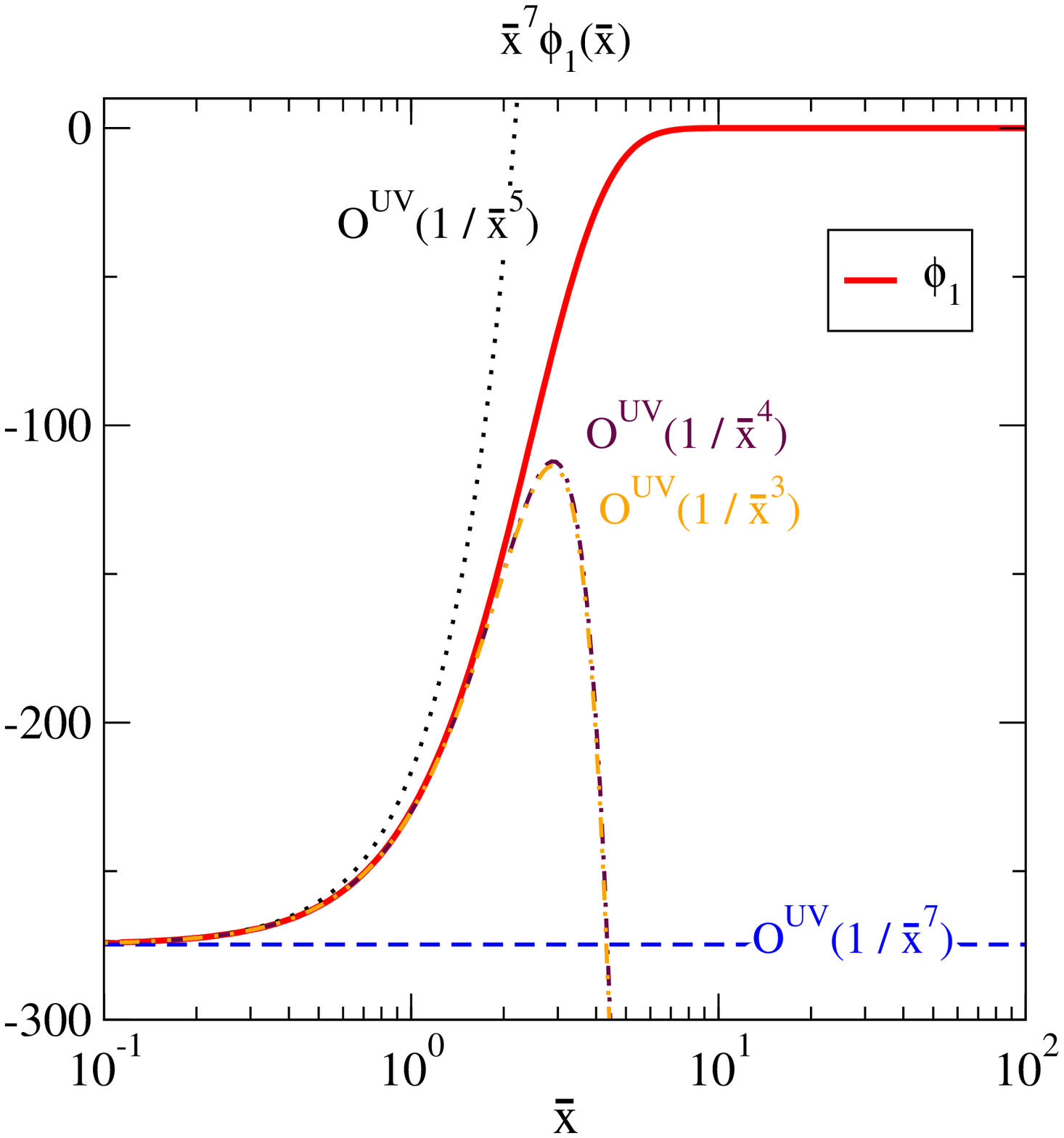}%
~~~\epsfysize=7.5cm\epsfbox{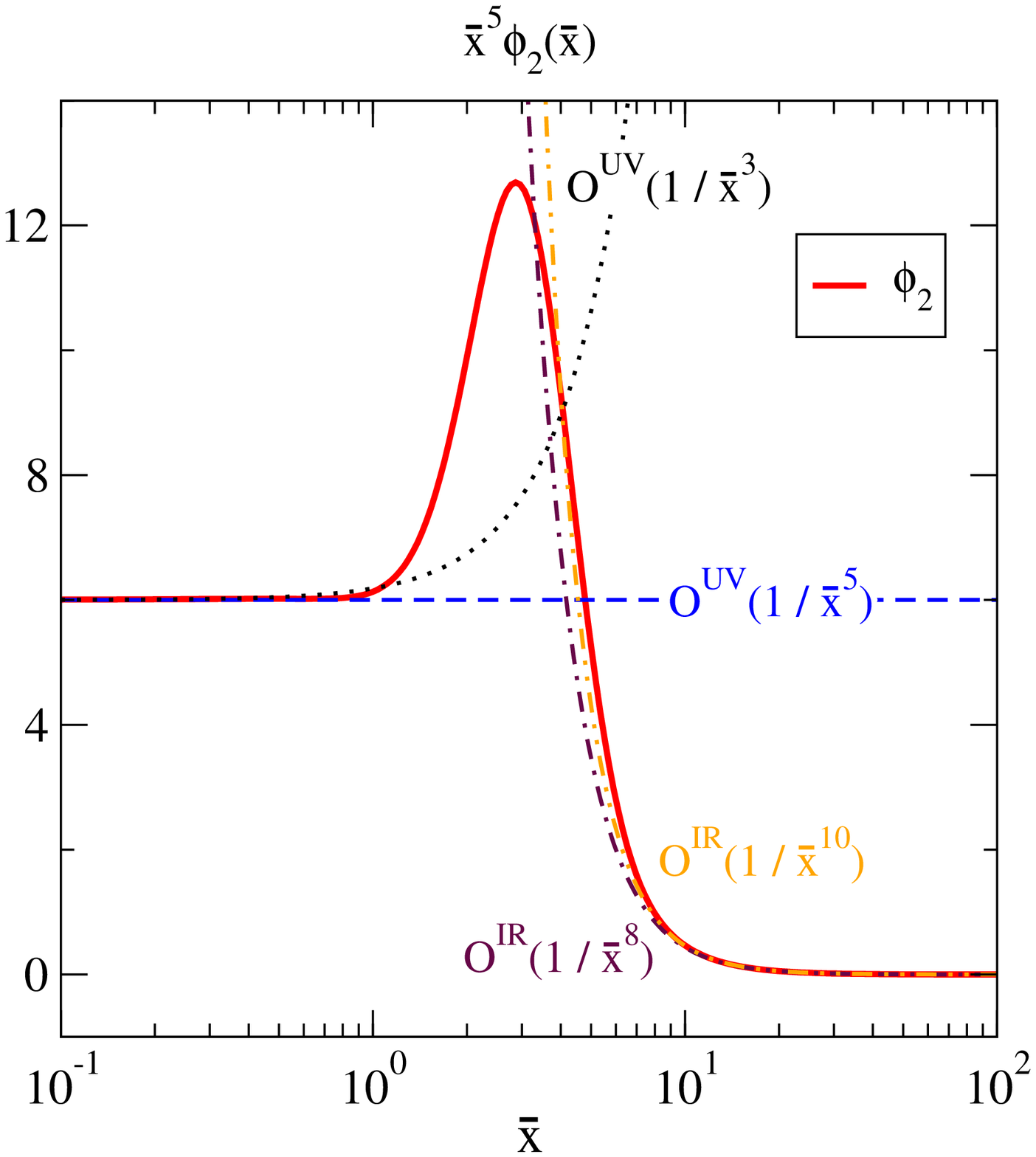}
}

\caption[a]{\small
The basic functions defined in 
\eqs\nr{f1_def}, \nr{f2_def}, together with their asymptotic
limits as given in \eqs\nr{f1_UV},
% \nr{f1_IR},
\nr{f2_UV},
\nr{f2_IR}.
The functions $O^\rmii{UV/IR}$ indicate 
the nature of the limit, and the argument shows the 
order up to which terms have been included. 
}
\la{fig:basic_b}
\end{figure}
%%%%%%%%%%%%%%%%%%%%%%%%%%%%%%%%%%%%%%%%%%%%%%%%%%%%%%%%%%%%%%%%%%%%%%%%%%%

Let us work out the asymptotic behaviours
of $\phi_1$ and $\phi_2$. For $\phi_1$, we find
\ba
 \phi_1^\rmii{UV}(\bar{x}) & = & 
 -\frac{824}{3\bar{x}^7} + \frac{58}{\bar{x}^5}-\frac{40}{3\bar{x}^4}
 -\frac{1}{45\bar{x}^3} + \rmO\biggl(\frac{1}{\bar{x}^2} \biggr) 
 \;, \quad \bar{x} \ll 1 
 \;, \la{f1_UV} \\ 
 \phi_1^\rmii{IR}(\bar{x}) & = & 
 \rmO(e^{-2\bar{x}})
 \;, \quad \bar{x} \gg 1
 \;. \la{f1_IR} 
\ea
For $\phi_2$, in turn, the short-distance behaviour is 
\be
 \phi_2^\rmii{UV}(\bar{x}) = \frac{6}{\bar{x}^5} + \frac{5}{27\bar{x}^3} + 
 \rmO\biggl( \frac{1}{\bar{x}} \biggr)
 \;, \quad \bar{x} \ll 1 
 \;. \la{f2_UV} 
\ee
The infrared behaviour is, in contrast, rather non-trivial: 
while all terms in $\phi_1$ are exponentially small at long 
distances, there are some parts in $\phi_2$ which are not. 
Indeed, despite a large $\bar{x}$, there is a slice
in the integration range of \eq\nr{f2_def} where  
$b$ is small
(because of the appearance of the unit vectors, this is 
not possible in the terms involving $a$ or $c$). 
The intuitive meaning of this term is perhaps clearest
if it is rewritten in momentum space: 
\ba 
 \phi_2^\rmii{IR}(\bar{x}) & \equiv &
 \frac{1}{2\pi\bar{x}}
    \int_0^\infty \!\! \dd \tilde{y}\: 
    \dd \tilde{z} \int_{-1}^1 \!\!  
    \dd t_1\: \dd t_2 \int_0^{2\pi} \!\!\dd\phi\;
    \frac{\tilde{y}\tilde{z}}
   {|\mathbf{e-\tilde{y}}|
    |\mathbf{e-\tilde{z}}|
    |\mathbf{\tilde{y}-\tilde{z}}|}
   \frac{\dd^6}{\dd \bar{x}^6} 
   \Bigg[\frac{\bar{x}^2}
   {e^{\bar{x} (\tilde{y} + \tilde{z} + 
    |\mathbf{\tilde{y}-\tilde{z}}| ) } - 1} 
   \Bigg] 
 \nn 
 & = &
 \frac{4\, \nabla^6}{\pi^2 T^6}
 \int_\vec{p} \frac{e^{i\mathbf{ p \cdot x}}}{p^2}  
 \int_\vec{q} \frac{e^{i\mathbf{ q \cdot x}}}{q^2}
 \Tint{R}'
 \frac{1}{R^2 (q-R)^2(p+R)^2}
 \;.   \la{f2_IR_pre}
\ea
This corresponds to a contribution from graph (vii) 
of \fig\ref{fig:graphs} where a non-zero Matsubara
mode circulates in one of the triangles;  
even if we shrink this ``heavy-mode contribution''   
to a series of higher-order operators, there still 
remains an infrared contribution in which the remaining
two propagators have Matsubara zero modes. 
These power-law contributions can be obtained by 
expanding the result of the sum-integral over $R$ 
in powers of $p$ and $q$ and omitting contact terms, 
which immediately leads to 
\be
 \phi_2^\rmi{IR}(\bar{x}) = 
 \frac{360\zeta(3)}{\bar{x}^8} + 
 \frac{2520\zeta(5)}{\bar{x}^{10}} + 
 \rmO\biggl( \frac{1}{\bar{x}^{12}} \biggr)
 \;, \quad \bar{x} \gg 1 \;. 
 \la{f2_IR}
\ee

%%%%%%%%%%%%%%%%%%%%%%%%%%% SUBSECTION %%%%%%%%%%%%%%%%%%%%%%%%%%%%%%%
%
\subsection{Asymptotic behaviours}

Inserting the asymptotic expansions in \eqs\nr{fLO_asy},
\nr{fNLO_asy},
\nr{f1_UV},
\nr{f1_IR},
\nr{f2_UV},
\nr{f2_IR} into \eqs\nr{restheta}, \nr{reschi}, 
we can write down the asymptotic expansions for 
the functions $\bar G_\theta^\rmi{F}$ and $\bar G_\chi^\rmi{F}$
at small and large $\bar{x}$. 
In the short-distance limit we get 
\ba
 \frac{\bar G^\rmi{F,UV}_\theta(x)}{4 d_A c_\theta^2} &=&  
 \fr{g^4\pi^4 T^7}{\bar{x}^7}
 \biggl\{480+\fr{g^2\Nc}{\pi^2} 
 \biggl[ 
   440\ln\(e^{\gammaE}\bar{\Lambda}x\) -\fr{824}{3}
 \biggr] \biggr\}\nn
& & \; + \, 
 \fr{22\,g^6\Nc\pi^2T^7}{135\bar{x}^3}
 +{\mathcal O}\(\fr{g^4}{\bar{x}},\fr{g^6}{\bar{x}},\fr{g^8}{\bar{x}^7}\)
 \;, \quad \bar{x} \ll 1  \;,
 \label{UVtheta}\\
%%%%%
 \frac{\bar G^\rmi{F,UV}_\chi(x)}{-16 d_A c_\chi^2} &=&
 \fr{g^4\pi^4 T^7}{\bar{x}^7}
 \biggl\{480+\fr{g^2\Nc}{\pi^2} 
 \biggl[ 
   440\ln\(e^{\gammaE}\bar{\Lambda}x\) -\fr{104}{3}
 \biggr] \biggr\}\nn
& & \; + \, 
 \fr{22\,g^6\Nc\pi^2T^7}{135\bar{x}^3}
 +{\mathcal O}\(\fr{g^4}{\bar{x}},\fr{g^6}{\bar{x}},\fr{g^8}{\bar{x}^7}\)
 \;, \quad \bar{x} \ll 1 \;. \label{UVchi}
\ea
Note that the dominant terms, i.e.\ the first 
rows, are temperature-independent.
The expressions obtained agree with the OPE results that can be extracted from 
ref.~\cite{lvv}, but two features are worth stressing. First of all, 
the leading thermal correction is of $\rmO(g^6)$ and positive, 
whereas $\rmO(g^4)$ only gives a subleading 
thermal contribution at short distances; 
in equal-time correlators, in contrast, both orders contribute
to the leading term $\rmO(-1/\bar{x}^4)$, 
which is negative at short distances~\cite{Iqbal:2009xz,lvv}. 
This drastic
difference illustrates that averaging over the Euclidean time coordinate
has a significant effect if $x \lsim \beta$. Second, the OPE contributions
proportional to $(e-3p)(T)$, which come with opposite signs in the two 
channels and distinguish them from each other~\cite{ope,lvv}, 
are of $\rmO(g^8/\bar{x}^3)$ and thus beyond the accuracy
of our present $\rmO(g^6)$ analysis. 
 
% Note that dominant terms, i.e.\ the first 
% rows, are temperature-independent.

At long distances, on the other hand, we obtain
\ba
 \frac{\bar G^\rmi{F,IR}_\theta(x)}{4 d_A c_\theta^2} &=&  
 g^4\pi^4 T^7 \biggl\{ \fr{96}{\bar{x}^6}
 +\frac{g^2\Nc}{\pi^2} 
 \biggl[ -\fr{4}{3\bar{x}^4}
 +\fr{24}{\bar{x}^5}
 +\fr{88}{\bar{x}^6} \ln \frac{e^{\gammaE}\bar{\Lambda}}{ 4\pi T } 
 \biggr]\biggr\}
 \nn 
 & & \; + \,  
 %\biggl\{ 
   \fr{360\zeta(3) g^6 \Nc \pi^2 T^7}{\bar{x}^8}
 %  +\fr{2520\zeta(5)}{\bar{x}^{10}}
 % \biggr\}
  + \rmO \biggl(g^4 e^{-2\bar{x}}, \fr{g^6}{\bar{x}^{10}}, 
  \frac{g^8}{\bar{x}^2} \biggr)
 \;, \quad \bar{x} \gg 1 
 \;,  \label{IRftheta}\\
%%%%%%
 \frac{\bar G^\rmi{F,IR}_\chi(x)}{-16 d_A c_\chi^2} &=&
  g^4\pi^4 T^7 \biggl\{ \fr{96}{\bar{x}^6}
 +\frac{g^2\Nc}{\pi^2} 
 \biggl[ -\fr{4}{3\bar{x}^4}
 +\fr{24}{\bar{x}^5}
 +\fr{88}{\bar{x}^6} 
 \biggl( \ln \frac{e^{\gammaE}\bar{\Lambda}}{ 4\pi T }  
   + \frac{6}{11} \biggr)
 \biggr]\biggr\}
 \nn 
 & & \; + \,  
 %\biggl\{ 
   \fr{360\zeta(3) g^6 \Nc \pi^2 T^7}{\bar{x}^8}
 %  +\fr{2520\zeta(5)}{\bar{x}^{10}}
 % \biggr\}
  + \rmO \biggl(g^4 e^{-2\bar{x}}, \fr{g^6}{\bar{x}^{10}}, 
  \frac{g^8}{\bar{x}^2} \biggr)
 \;, \quad \bar{x} \gg 1 
 \;. \label{IRfchi}
\ea
The first rows indicate
that at sufficiently large distances, 
$
 \bar{x}^2 \gg \pi^2/g^2\Nc
$,
the correction of $\rmO(g^6)$ overtakes the term of $\rmO(g^4)$; 
this signals a breakdown of the perturbative series and 
necessitates a resummation, to which we now turn. 

%%%%%%%%%%%%%%%%%%%%%%%%%%%%% SECTION %%%%%%%%%%%%%%%%%%%%%%%%%%%%%%%
%
\section{Effective theory computation}
\la{se:ir}

As can be observed from \eqs\nr{IRftheta}, \nr{IRfchi}, 
at distances of the order $x^2 \sim 1/g^2T^2 \Nc$ the naive
perturbative series breaks down and needs to be resummed in order 
to  recover the correct long-distance behaviour. 
A systematic way to implement these resummations
is provided by effective theories. 
In our case the relevant low-energy effective theory is known as EQCD,
a three-dimensional theory for the Matsubara zero  modes 
of $A_i^a$ and $A_0^a$~\cite{pg,ap}.

Keeping the dimensions of the fields as they are in the original
theory, and continuing to denote by $g^2$ the renormalized 4d gauge
coupling, the operators we are interested in can be represented 
in EQCD as
\begin{align}
 \theta^\rmi{E} &= c_\theta\, g^2 \left[
        \mathcal{Z}_\rmi{F} \tilde F_{ij}^a \tilde F_{ij}^a
    + 2 \mathcal{Z}_\rmi{D} (\mathcal{D}_i \tilde A_0)^a 
   (\mathcal{D}_i \tilde A_0)^a
    +   \ldots \right] \;, \\
 \chi^\rmi{E} &= c_\chi\, g^2 \left[ -4 \mathcal{Z}_\rmi{E} 
 \,\epsilon_{ijk}\, (\mathcal{D}_i \tilde A_0)^a \tilde F_{jk}^a 
    + \ldots
 \right]
 \;,
\end{align}
where the numerical coefficients have been chosen so  
that the renormalization factors 
$\mathcal{Z}_\rmi{F}$, $\mathcal{Z}_\rmi{D}$, $\mathcal{Z}_\rmi{E}$
are all of the form $1+\mathcal{O}(g^2)$.
Dropping again contact terms, and setting $\epsilon=0$ since 
the contributions are ultraviolet finite, the leading-order
results can be written as   
\begin{align}
 \frac{\bar G^\rmi{E}_\theta(x)}{4 d_A c_\theta^2 g^4} &=
    T \int_\vec{p} e^{i\mathbf{p \cdot x}} 
     \int_\vec{q} \biggl[ 
        \frac{\mathcal{Z}_\rmi{F}^2}{2} \frac{p^4}{q^2(p-q)^2}
    +
        \frac{\mathcal{Z}_\rmi{D}^2}{2} 
        \frac{(p^2+2\mE^2)^2}{(q^2+\mE^2)[(p-q)^2+\mE^2]}
	\biggr]
 \nn 
 &= 
 \frac{\mathcal{Z}_\rmi{F}^2 T}{2(4\pi)^2 x} \, \frac{{\rm d}^4}{{\rm d} x^4} 
 \biggl( \frac{1}{x} \biggr)
 + \frac{\mathcal{Z}_\rmi{D}^2 T}{2(4\pi)^2  x} \, 
 \Bigl(\frac{{\rm d}^2}{{\rm d} x^2} -2\mE^2\Bigr)^2\, 
 \biggl( \frac{e^{-2\mE x} }{x} \biggr)
 \;, \label{eq:theta_eqcd} \\
 \frac{\bar G^\rmi{E}_\chi(x)}{-16 d_A c_\chi^2 g^4 } &=
   T \int_\vec{p} e^{i\mathbf{p \cdot x}} 
    \int_\vec{q} \frac{\mathcal{Z}_\rmi{E}^2 (p^2+\mE^2)^2}
    {(q^2+\mE^2)(p-q)^2} 
 \nn
    &= \frac{\mathcal{Z}_\rmi{E}^2 T}{(4\pi)^2x}\, 
  \Bigl(\frac{{\rm d}^2}{{\rm d} x^2} -\mE^2 \Bigr)^2\, 
  \biggl( \frac{e^{-\mE x}}{x}  \biggr)
 \;, \label{eq:chi_eqcd}
\end{align}
where $\mE^2$ is the Debye mass parameter appearing in the EQCD Lagrangian. 
To avoid double-counting, we write the observables as
\begin{equation}
 \bar{G}_{\theta,\chi}(x) = \Big[ 
 \bar{G}^\rmi{F}_{\theta,\chi}(x)
 - \bar{G}_{\theta,\chi}^\rmi{E}(x) \Big]_\rmii{naive}
 + \Big[ \bar{G}_{\theta,\chi}^\rmi{E}(x)\Big]_\rmii{resummed}
 \; ,
 \label{eq:resum_recipe}
\end{equation}
where the quantity in the first square brackets is infrared safe. 
In the ``naive'' results, $\mE$ is treated as a perturbatively 
small quantity of $\rmO(g)$, and we can expand with respect 
to it in eqs.~(\ref{eq:theta_eqcd}), (\ref{eq:chi_eqcd}), giving
\ba
  \biggl( \frac{\bar G^\rmi{E}_\theta(x)}{4 d_A c_\theta^2 g^4}
  \biggr)_\rmii{naive} & = & 
  \mathcal{Z}_\rmi{F}^2 \biggl[ \frac{48 \pi^4 T^7 }{\bar{x}^6} \biggr]
  + 
  \mathcal{Z}_\rmi{D}^2 \biggl[ \frac{48 \pi^4 T^7 }{\bar{x}^6}
  - \frac{4 \pi^2 T^5 \mE^2}{\bar{x}^4} + \rmO\Bigl(\mE^4\Bigr) \biggr]
 \;, \\ 
  \biggl( \frac{\bar G^\rmi{E}_\chi(x)}{-16 d_A c_\chi^2 g^4}
  \biggr)_\rmii{naive} & = & 
  \mathcal{Z}_\rmi{E}^2 \biggl[ \frac{96 \pi^4 T^7 }{\bar{x}^6}
  - \frac{4 \pi^2 T^5 \mE^2}{\bar{x}^4} + \rmO\Bigl(\mE^4\Bigr) \biggr]
 \;.  \la{eqcd_chi_naive}
\ea
Substituting $\mE^2 = g^2 T^2 \Nc/3$,  
$\mathcal{Z}_\rmi{F} = \mathcal{Z}_\rmi{D} = \mathcal{Z}_\rmi{E} =1$ 
and comparing with \eqs\nr{IRftheta} and \nr{IRfchi}, we 
see that these reproduce the dominant LO and NLO
long-distance behaviours. Therefore, 
in both channels, the subtraction-addition step 
of \eq\nr{eq:resum_recipe} replaces the leading infrared
contribution $\sim - g^6 \Nc/\bar{x}^4$ by an exponentially suppressed 
behaviour, proportional to $\sim \exp(-\mE x)$ in the pseudoscalar
($\chi$) and to $\sim \exp(- 2 \mE x)$ in the  scalar ($\theta$) channel, 
thereby postponing the breakdown of the perturbative series to larger
distances. 

As far as the subleading infrared terms are concerned, we would 
expect those of $\rmO(g^6 \Nc/\bar{x}^5)$ to also be replaced by 
exponentially suppressed terms in the pseudoscalar channel ($\chi$), 
in which symmetries require the appearance of an odd number of 
$\tilde A_0^a$ propagators in any intermediate state~\cite{Arnold:1995bh}.
(In other words, the effective theory
for the static colour-magnetic modes, MQCD, does not contain 
operators with the right quantum numbers to contribute to 
$\bar{G}_\chi$.)
In contrast, in the scalar channel, power-law terms of 
$\rmO(g^6 \Nc/\bar{x}^5)$ remain over, and imply that our computation
becomes unreliable at the distance scale where this term overtakes
the leading-order term, 
i.e.\ when $\rmO(1/\bar{x}^6)\sim \rmO(g^2\Nc/\pi^2\bar{x}^5)$. This happens
at the non-perturbative colour-magnetic scale $x \sim \pi/(g^2\Nc T)$;
the corresponding behaviour could in principle be determined 
through a non-perturbative analysis of MQCD.

As far as the remaining power-law terms go, those of 
$\rmO(g^6\Nc T^7/\bar{x}^6)$ could be accounted for by a choice
of the $\rmO(g^2)$ corrections to the renormalization factors 
$\mathcal{Z}_\rmi{F}, \mathcal{Z}_\rmi{D}, \mathcal{Z}_\rmi{E}$.
Unfortunately, our present results do not suffice to fix all of 
these unambigously,
but further matching computations would be 
needed. Higher powers still, $\rmO(g^6 T^7/\bar{x}^8)$ etc, 
correspond to higher-dimensional operators within EQCD or MQCD.
Since the focus of the present paper is on intermediate distances, 
we have not carried out a next-to-leading order analysis of the 
correlators on the EQCD side. Rather, we will 
use leading-order resummation, as implied 
by \eqs\nr{eq:theta_eqcd}--\nr{eqcd_chi_naive}, to gauge 
the distance scale where infrared effects become important, 
and declare our results as unreliable above this scale.  
Explicitly, our resummation amounts to the additional contributions
\ba
 \frac{\delta_\rmii{resum} \bar{G}_\theta(x)}
 {4 d_A c_\theta^2} & = & 
 -g^4 \pi^4 T^7 \biggl[ \frac{48}{\bar{x}^6} 
 - \frac{4 g^2 \Nc}{3 \pi^2 \bar{x}^4}
  \biggr] + 
  \frac{g^4 \mE^6 T}{32 \pi^2 \tilde{x}} \biggl(
  \frac{{\rm d}^2 }{{\rm d}\tilde{x}^2} -2 \biggr)^2 
  \biggl( \frac{e^{-2\tilde x}}{\tilde{x}} \biggr)
 \;, \la{theta_resum}  \\ 
 \frac{\delta_\rmii{resum} \bar{G}_\chi(x)}
 {-16 d_A c_\chi^2} & = & 
 -g^4 \pi^4 T^7 \biggl[ \frac{96}{\bar{x}^6} 
 - \frac{4 g^2 \Nc}{3 \pi^2 \bar{x}^4}
  \biggr] + 
  \frac{g^4 \mE^6 T}{16 \pi^2 \tilde{x}} \biggl(
  \frac{{\rm d}^2 }{{\rm d}\tilde{x}^2} -1 \biggr)^2 
  \biggl( \frac{e^{-\tilde x}}{\tilde{x}} \biggr)
 \;, \la{chi_resum} 
\ea
to be added to 
\eqs\nr{restheta}, \nr{reschi}, with $\tilde x \equiv \mE x$.

%%%%%%%%%%%%%%%%%%%%%%%%%%%% SECTION %%%%%%%%%%%%%%%%%%%%%%%%%%%%%%%
%
\section{Numerical evaluation}
\la{se:plot}

\subsection{Choice of the renormalization scale}

Upon adding together the full theory and EQCD contributions 
to the correlators, \eqs\nr{restheta} and \nr{theta_resum} 
for the $\theta$ channel and \eqs\nr{reschi} and \nr{chi_resum} 
for the $\chi$ channel, 
we are now ready to numerically estimate the behaviour of 
the functions in various regimes. We do this for a variety 
of distance scales, ranging from $x\ll 1/\pi T$ to $x\gg 1/\pi T$, 
which in particular implies that we should choose 
an ``optimal'' renormalization scale that adapts to the physical quantity 
and distance regime of interest. 
At short distances, it is natural to determine 
the optimal scale from the UV limit of the correlators, 
demanding that the leading $\bar{\Lambda}$-dependent term in 
the NLO part of the results vanishes.
According to \eqs\nr{UVtheta} and \nr{UVchi}
this yields
\be
 \bar{\Lambda}^x_\theta = \fr{e^{103/165-\gammaE}}{x}
 \;, \quad
 \bar{\Lambda}^x_\chi = \fr{e^{13/165-\gammaE}}{x}
 \;. \la{Lamx}
\ee
At long distances, on the other hand, we fix the scale from 
the logarithmic terms in \eqs\nr{IRftheta}, \nr{IRfchi}, which
correspond to NLO corrections to the factors 
$\mathcal{Z}_\rmi{F}, \mathcal{Z}_\rmi{D}, \mathcal{Z}_\rmi{E}$
of \se\ref{se:ir}
and are not otherwise cancelled by the resummation. 
This gives 
\be
 \bar{\Lambda}^T_\theta = e^{-\gammaE}\,4\pi T   
 \;, \quad
 \bar{\Lambda}^T_\chi = e^{-6/11-\gammaE}\,4\pi T
 \;. \la{LamT}
\ee
(Remarkably, in both channels, the short-distance and long-distance
scale choices coincide at $\bar{x} = e^{103/165-\ln2} \approx 0.9334$.) 
At intermediate distances, 
we make the simple choice of defining an ``optimal'' scale through
\ba
 \bar{\Lambda}^\rmi{opt}& \equiv &
 % \mathop{\rm max}(\bar{\Lambda}^x,\bar{\Lambda}^T) 
 \sqrt{(\bar{\Lambda}^T)^2+(\bar{\Lambda}^x)^2}
 \;. \label{lambdaoptgen}
\ea
% If, however, the zero-temperature result is considered
% separately, then only the first rows of \eqs\nr{UVtheta} and \nr{UVchi}
% contribute, and we fix $\bar{\Lambda}^\rmi{opt}_{\theta,\chi}$ to the 
% values of \eq\nr{Lamx} at all distances. 
Subsequently we vary our $\bar{\Lambda}$ parameter 
by a factor of 2 around this scale.

As far as the gauge coupling is concerned, we solve it 
from the 2-loop renormalization group equation, and define
\be
   \Lambdamsbar \equiv \lim_{\bmu\to\infty}
   \bmu \Bigl( b_0 g^2 \Bigr) ^{-b_1/2 b_0^2}
   \exp \biggl( -\frac{1}{2 b_0 g^2}\biggr)
 \;.
\ee 
Where needed, the Debye mass parameter is evaluated from the 
leading-order 
expression $\mE^2 = g^2 T^2\Nc/3$ with the same $g^2$ inserted
as elsewhere.

%%%%%%%%%%%%%%%%%%%%%%%%%%%%%% SUBSECTION %%%%%%%%%%%%%%%%%%%%%%%%%%%%
%
\subsection{Plots}

%%%%%%%%%%%%%%%%%%%%%%%%%%%%%%%%% FIGURE %%%%%%%%%%%%%%%%%%%%%%%%%%%%%%%%%
\begin{figure}[t]

%\vspace*{-3cm}

\centerline{%
 \epsfysize=7.5cm\epsfbox{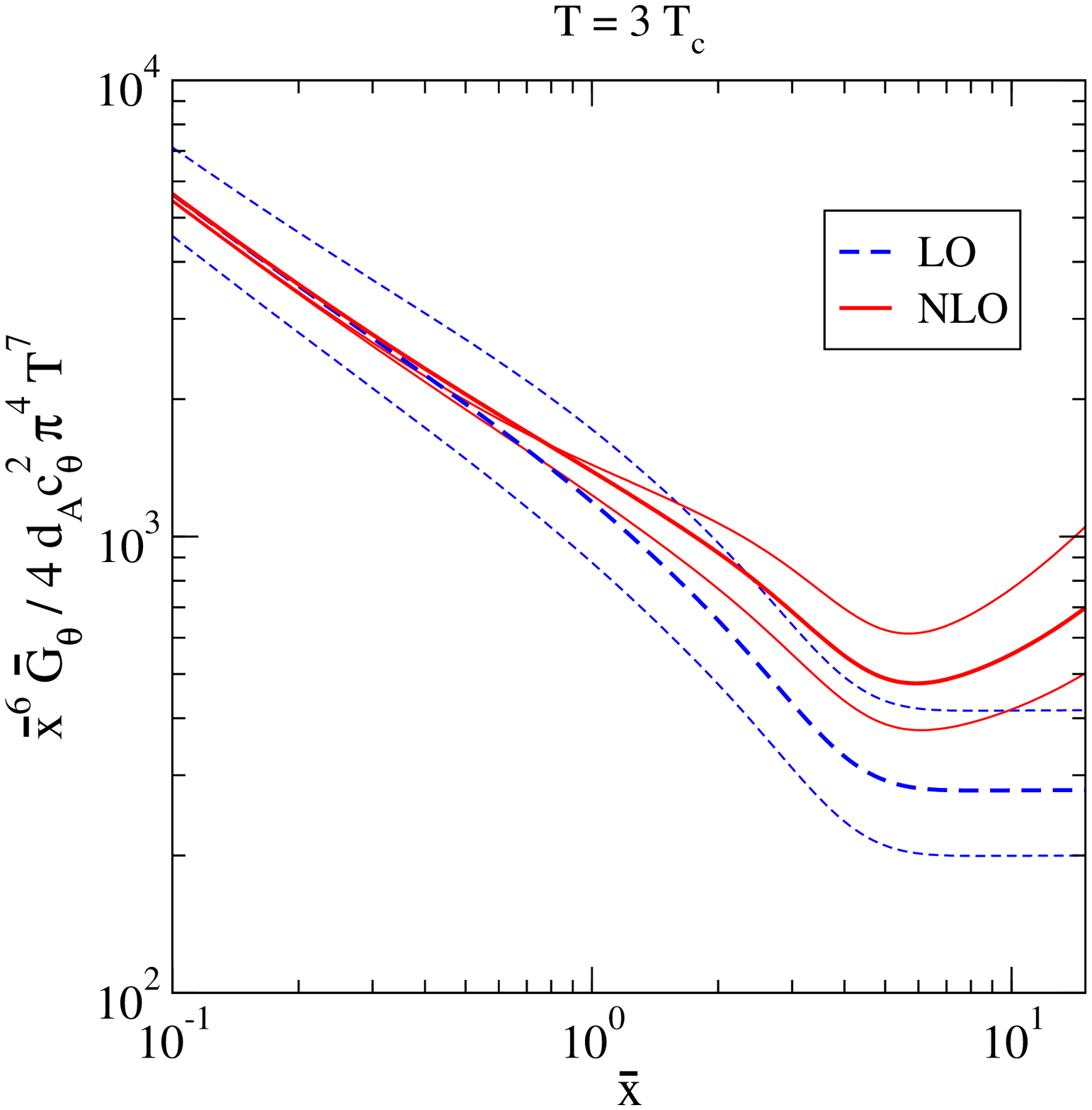}\;\;
 \epsfysize=7.5cm\epsfbox{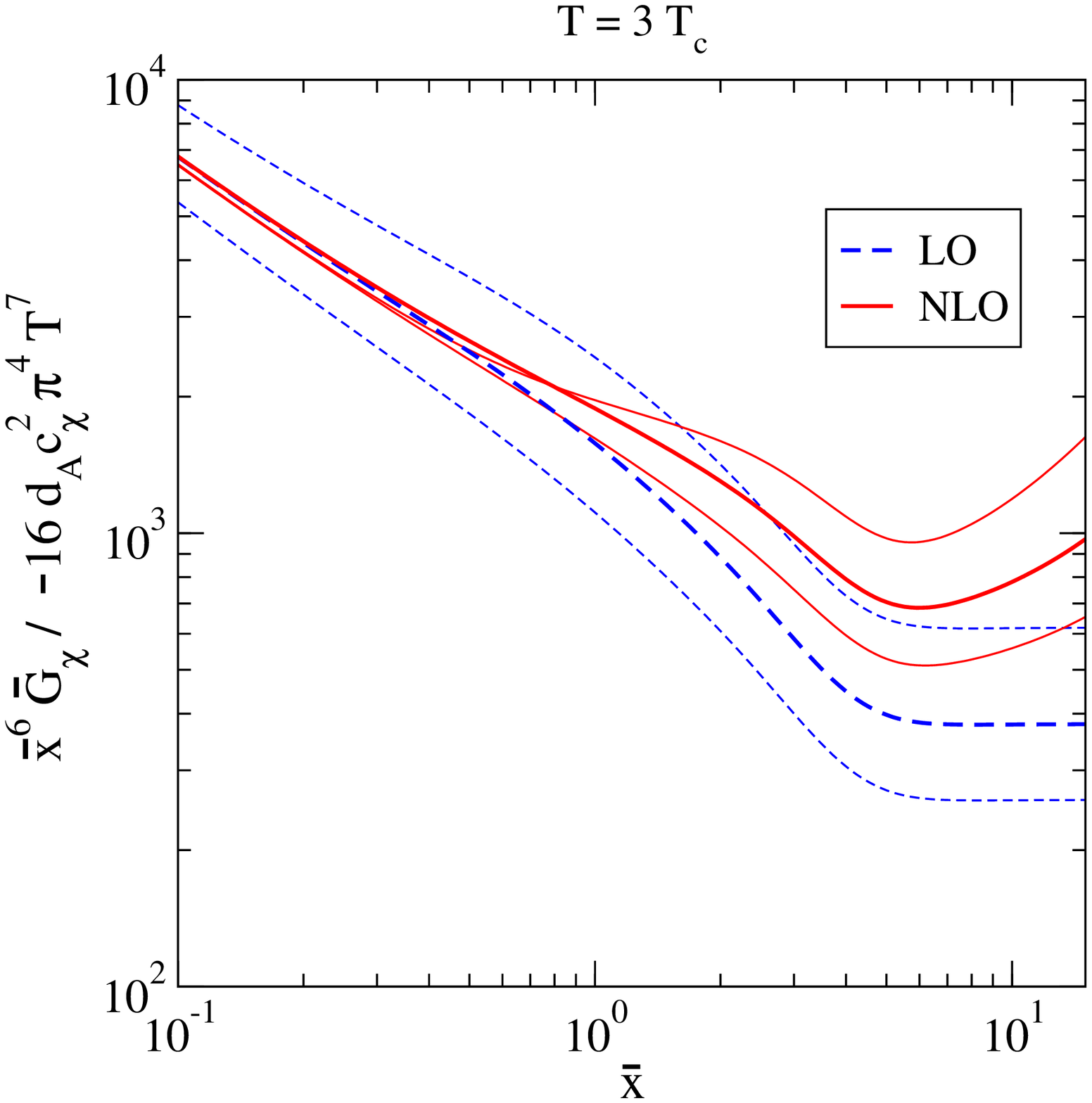}\;\;
}

\caption[a]{\small
The behaviour of the full $\theta$ (left) and $\chi$ (right) correlators,
multiplied by $\bar{x}^6$,  at $T=3\Tc$. Thick lines indicate 
the scale choice $\bar{\Lambda} = \bar{\Lambda}^\rmi{opt}$
(\eq\nr{lambdaoptgen}), thin lines variations by a factor of 2 
around this scale. Both correlators vary rapidly with $\bar{x}$
and are dominated by a vacuum-like behaviour up to 
distances $\bar{x} \sim $ a few. Our results become 
inaccurate at $\bar{x} \gsim 10$, cf.\ \fig\ref{fig:rdep5}.
 }
\la{fig:rdep1}
\end{figure}
%%%%%%%%%%%%%%%%%%%%%%%%%%%%%%%%%%%%%%%%%%%%%%%%%%%%%%%%%%%%%%%%%%%%%%%%%%%

%%%%%%%%%%%%%%%%%%%%%%%%%%%%%%%%% FIGURE %%%%%%%%%%%%%%%%%%%%%%%%%%%%%%%%%
\begin{figure}[t]

%\vspace*{-3cm}

\centerline{%
 \epsfysize=7.5cm\epsfbox{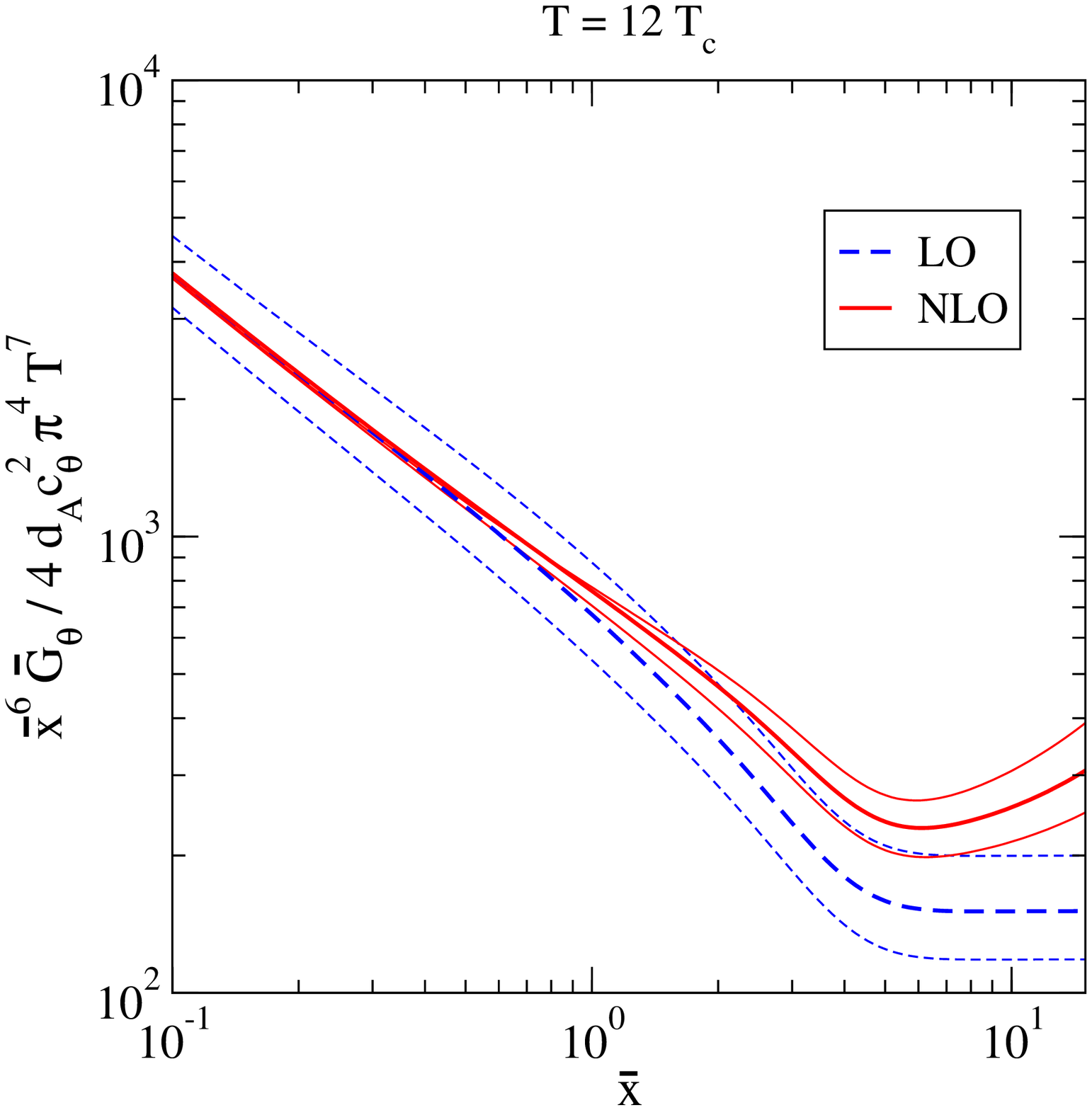}\;\;
 \epsfysize=7.5cm\epsfbox{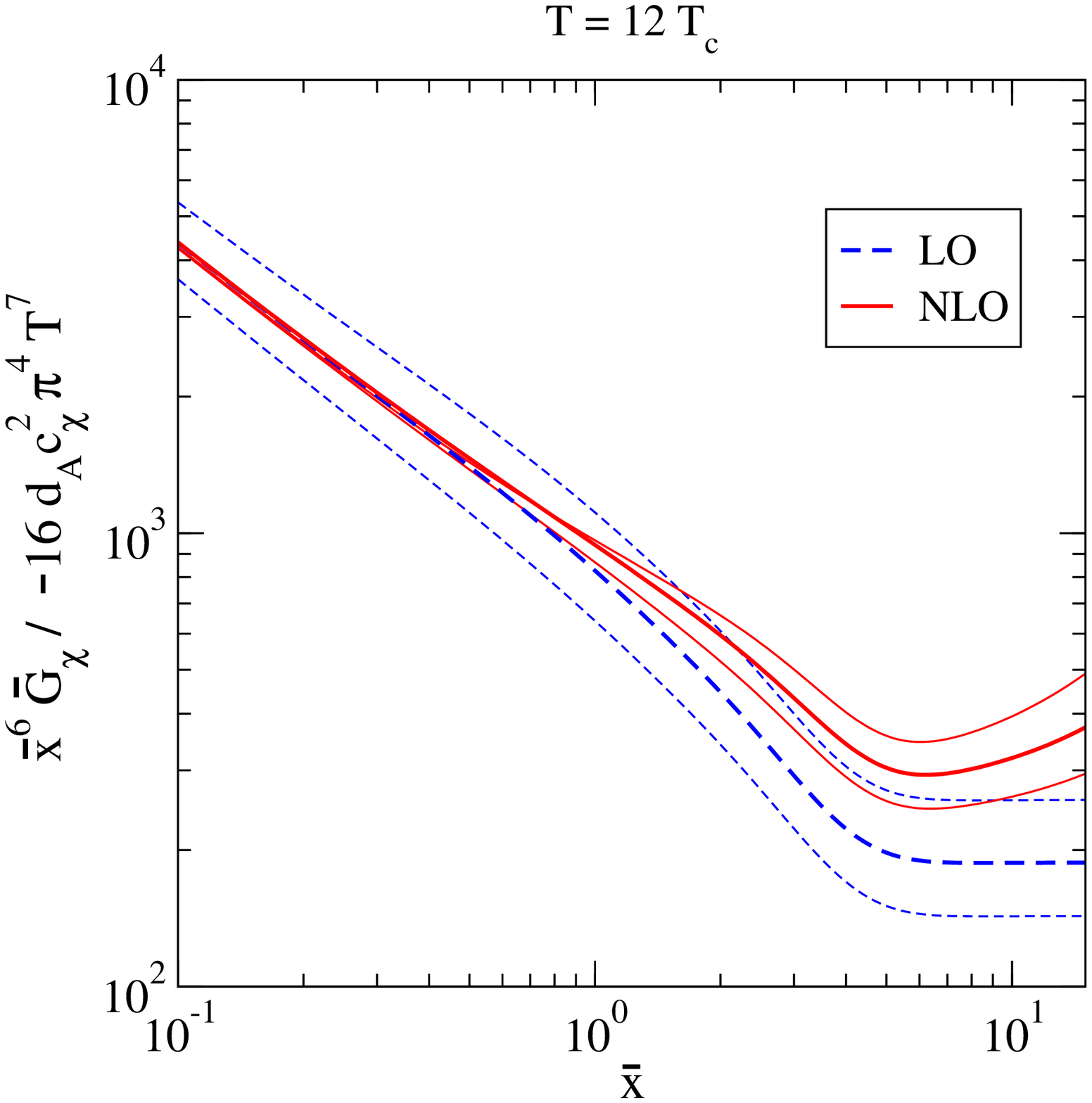}\;\;
}

\caption[a]{\small
Like \fig\ref{fig:rdep1} but at $T=12\Tc$. 
 }
\la{fig:rdep1_b}
\end{figure}
%%%%%%%%%%%%%%%%%%%%%%%%%%%%%%%%%%%%%%%%%%%%%%%%%%%%%%%%%%%%%%%%%%%%%%%%%%%

%%%%%%%%%%%%%%%%%%%%%%%%%%%%%%%%% FIGURE %%%%%%%%%%%%%%%%%%%%%%%%%%%%%%%%%
\begin{figure}[t]

%\vspace*{-3cm}

\centerline{%
 \epsfysize=7.5cm\epsfbox{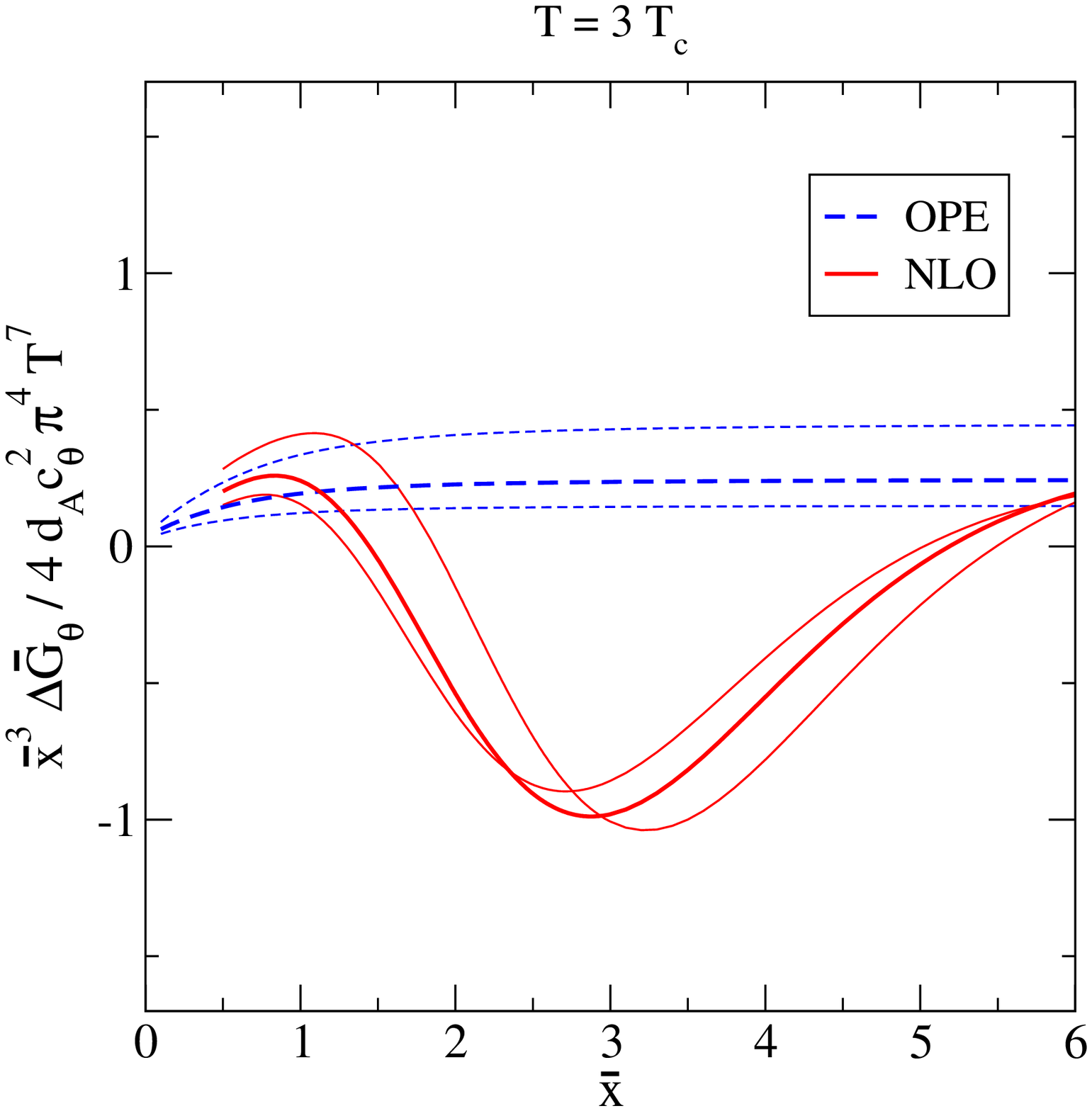}\;\;
 \epsfysize=7.5cm\epsfbox{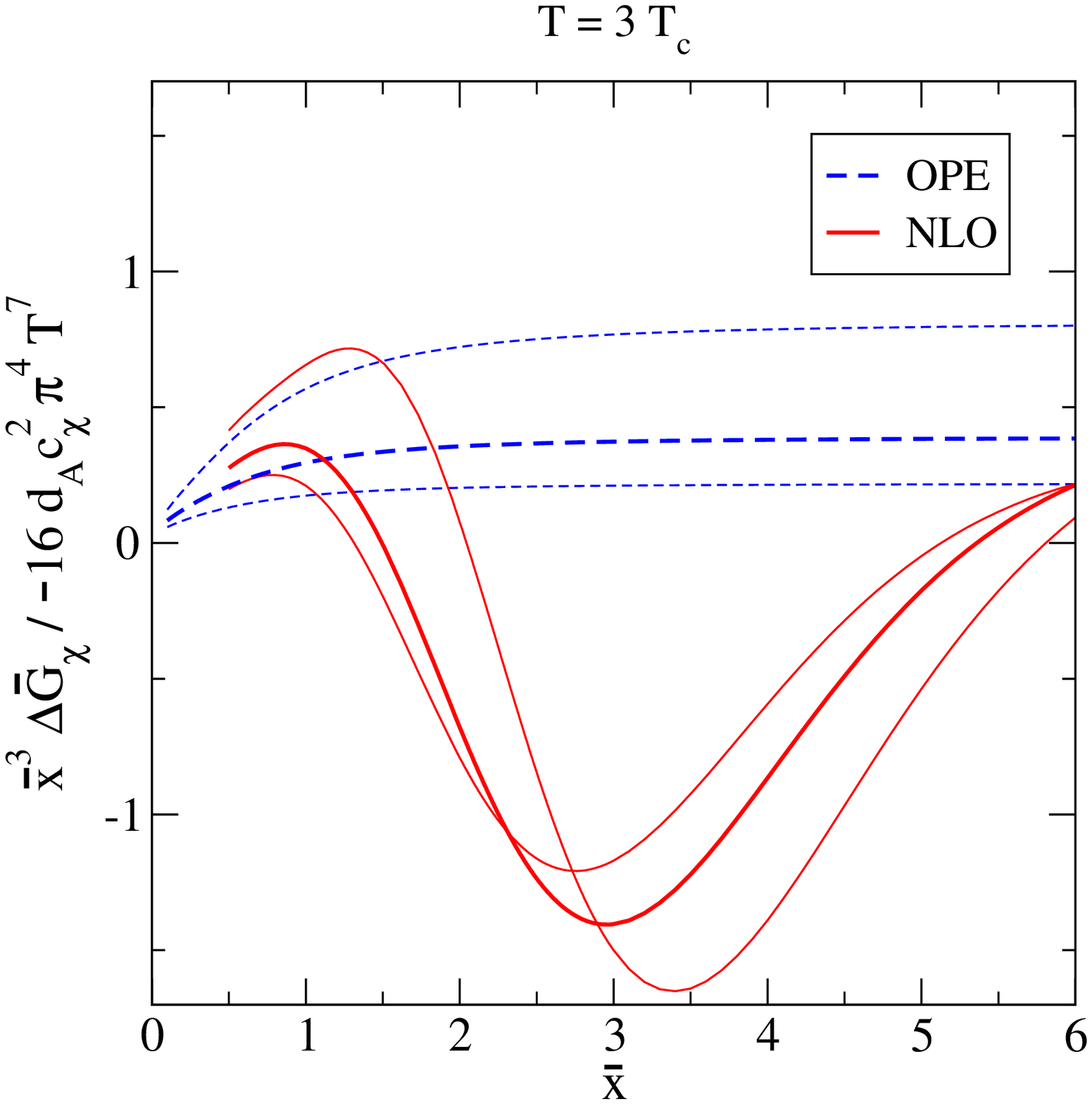}
}

\caption[a]{\small
The short-distance 
behaviours of the ``thermal'' parts of the $\theta$ (left) and $\chi$ (right) 
correlators, multiplied by $\bar{x}^3$, 
compared with the corresponding OPE limits 
(the second lines of \eqs\nr{UVtheta}, \nr{UVchi}), 
at $T = 3 \Tc$. Thick lines indicate 
the scale choice $\bar{\Lambda} = \bar{\Lambda}^\rmi{opt}$, thin 
lines variations by a factor of 2 around this scale. (Numerical evaluation
is challenging at $\bar{x} \ll 1$ due to significance loss:
not only the ``vacuum'' parts $1/\bar{x}^7$ but also ``thermal'' 
parts behaving as $1/\bar{x}^5$ and $1/\bar{x}^4$ cancel in 
$\Delta \bar{G}_{\theta,\chi}$ at $\bar{x}\lsim 1$,  
cf.\ \eqs\nr{fNLO_asy}, \nr{f1_UV}, \nr{f2_UV}).
 }
\la{fig:rdep4}
\end{figure}
%%%%%%%%%%%%%%%%%%%%%%%%%%%%%%%%%%%%%%%%%%%%%%%%%%%%%%%%%%%%%%%%%%%%%%%%%%%

%%%%%%%%%%%%%%%%%%%%%%%%%%%%%%%%% FIGURE %%%%%%%%%%%%%%%%%%%%%%%%%%%%%%%%%
\begin{figure}[t]

%\vspace*{-3cm}

\centerline{%
 \epsfysize=7.5cm\epsfbox{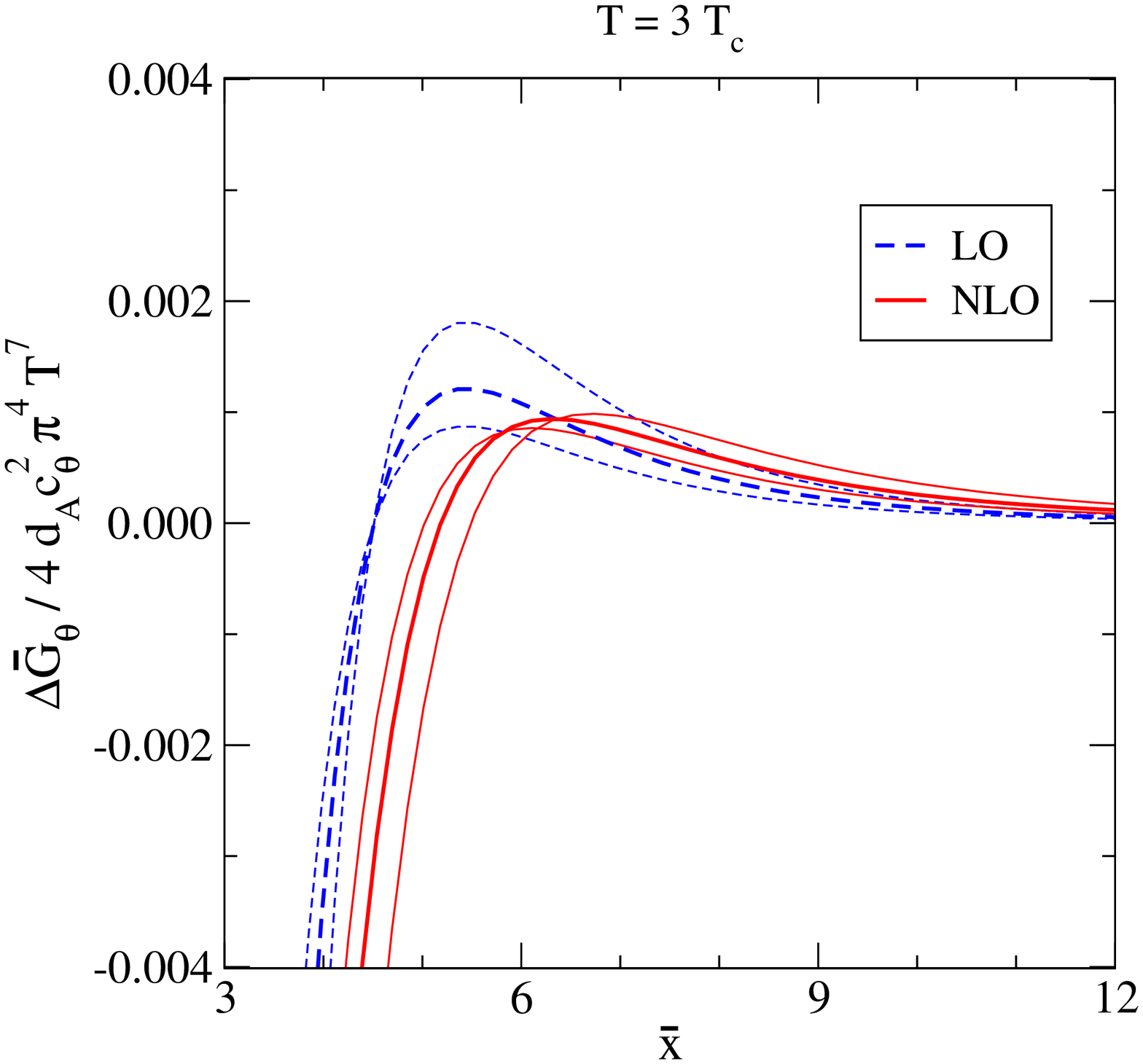}\;\;
 \epsfysize=7.5cm\epsfbox{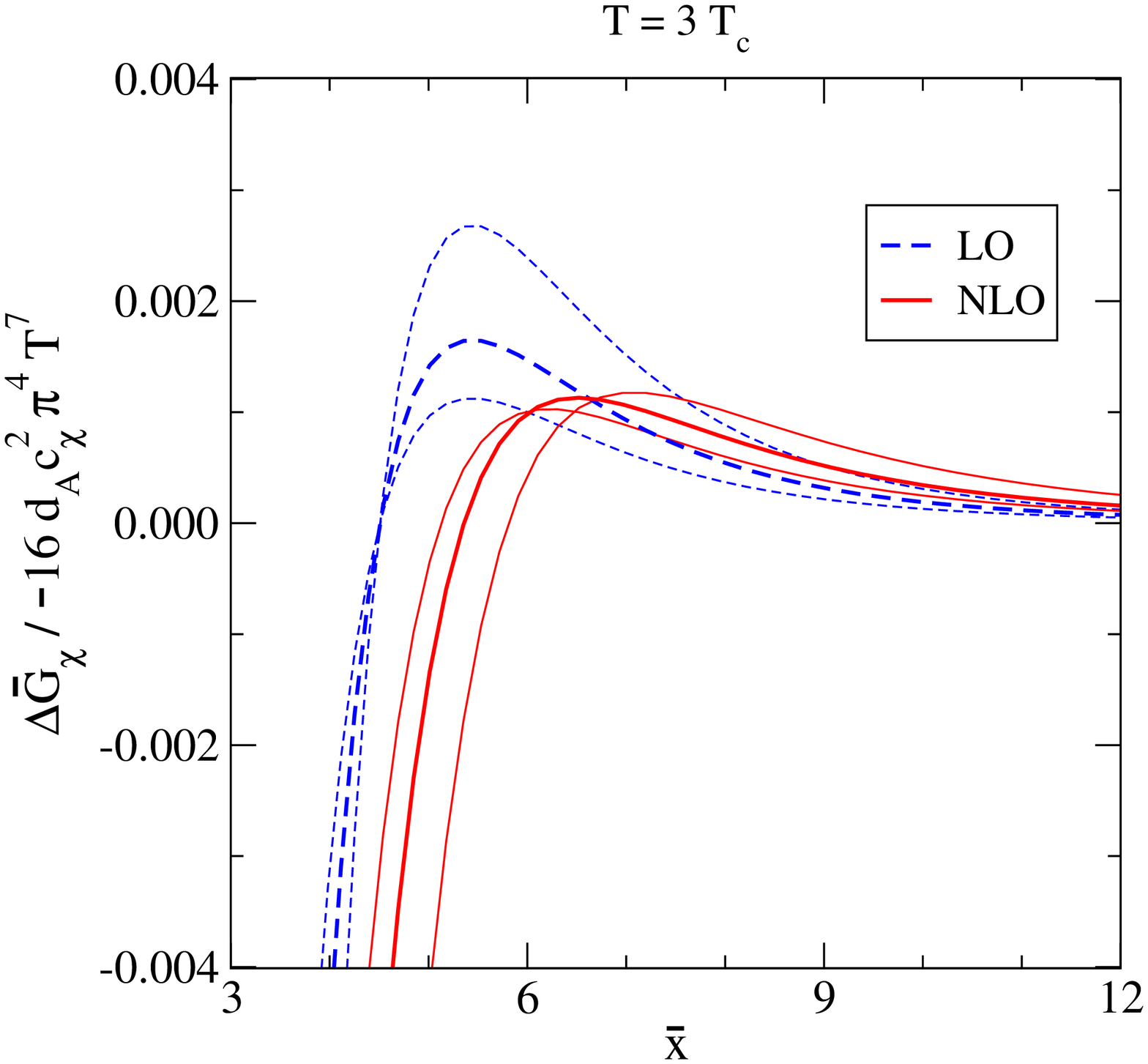}
}

\caption[a]{\small
The ``thermal'' parts of the $\theta$ (left) 
and $\chi$ (right) correlators at intermediate distances, 
at $T = 3 \Tc$. Thick lines indicate the scale choice 
$\bar{\Lambda} = \bar{\Lambda}^\rmi{opt}$, thin lines 
variations by a factor of 2 around this scale. Correlations
are slightly more pronounced in the $\chi$ channel. 
 }
\la{fig:rdep3}
\end{figure}
%%%%%%%%%%%%%%%%%%%%%%%%%%%%%%%%%%%%%%%%%%%%%%%%%%%%%%%%%%%%%%%%%%%%%%%%%%%

%%%%%%%%%%%%%%%%%%%%%%%%%%%%%%%%% FIGURE %%%%%%%%%%%%%%%%%%%%%%%%%%%%%%%%%
\begin{figure}[t]

%\vspace*{-3cm}

\centerline{%
 \epsfysize=7.5cm\epsfbox{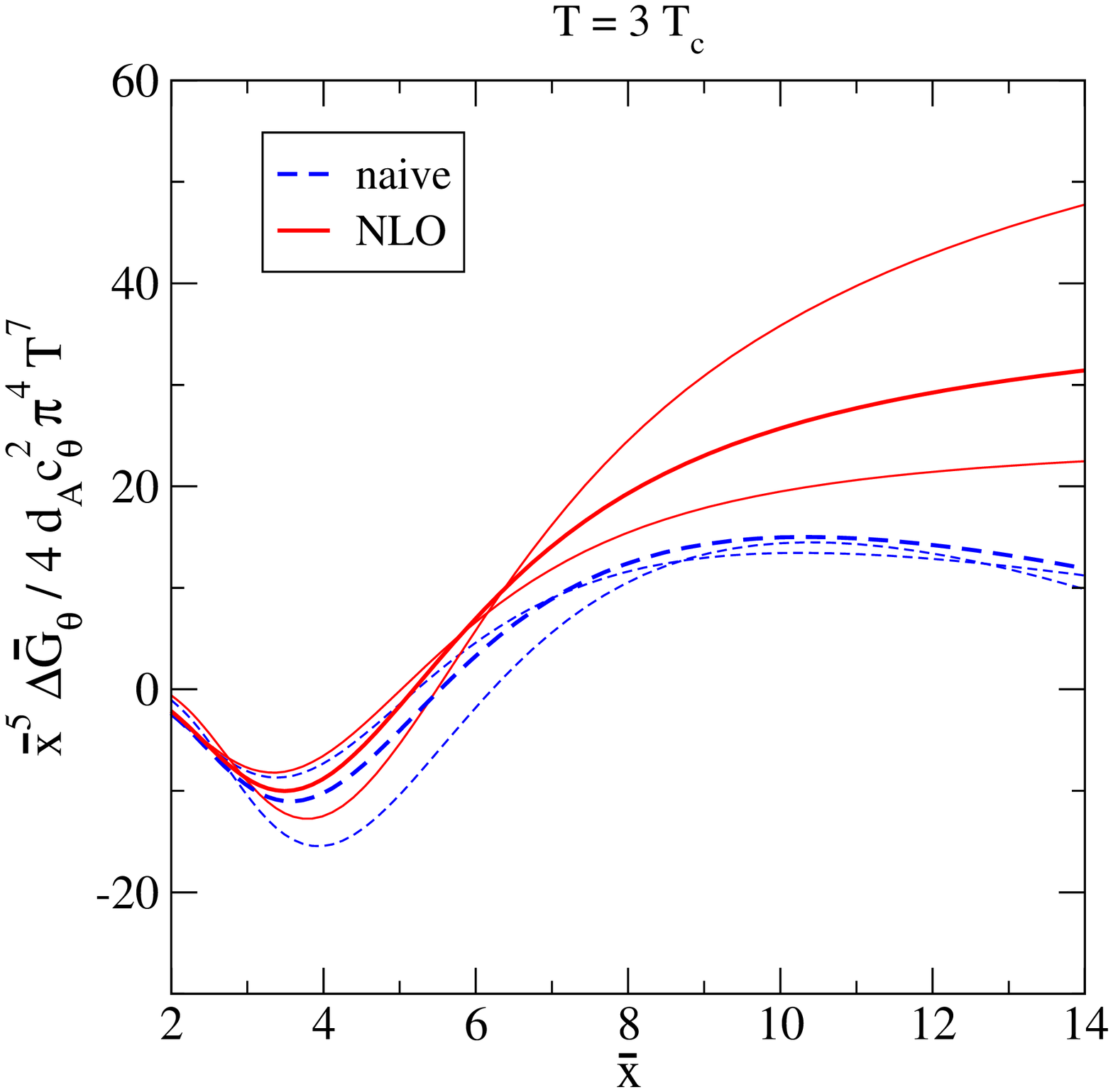}\;\;
 \epsfysize=7.5cm\epsfbox{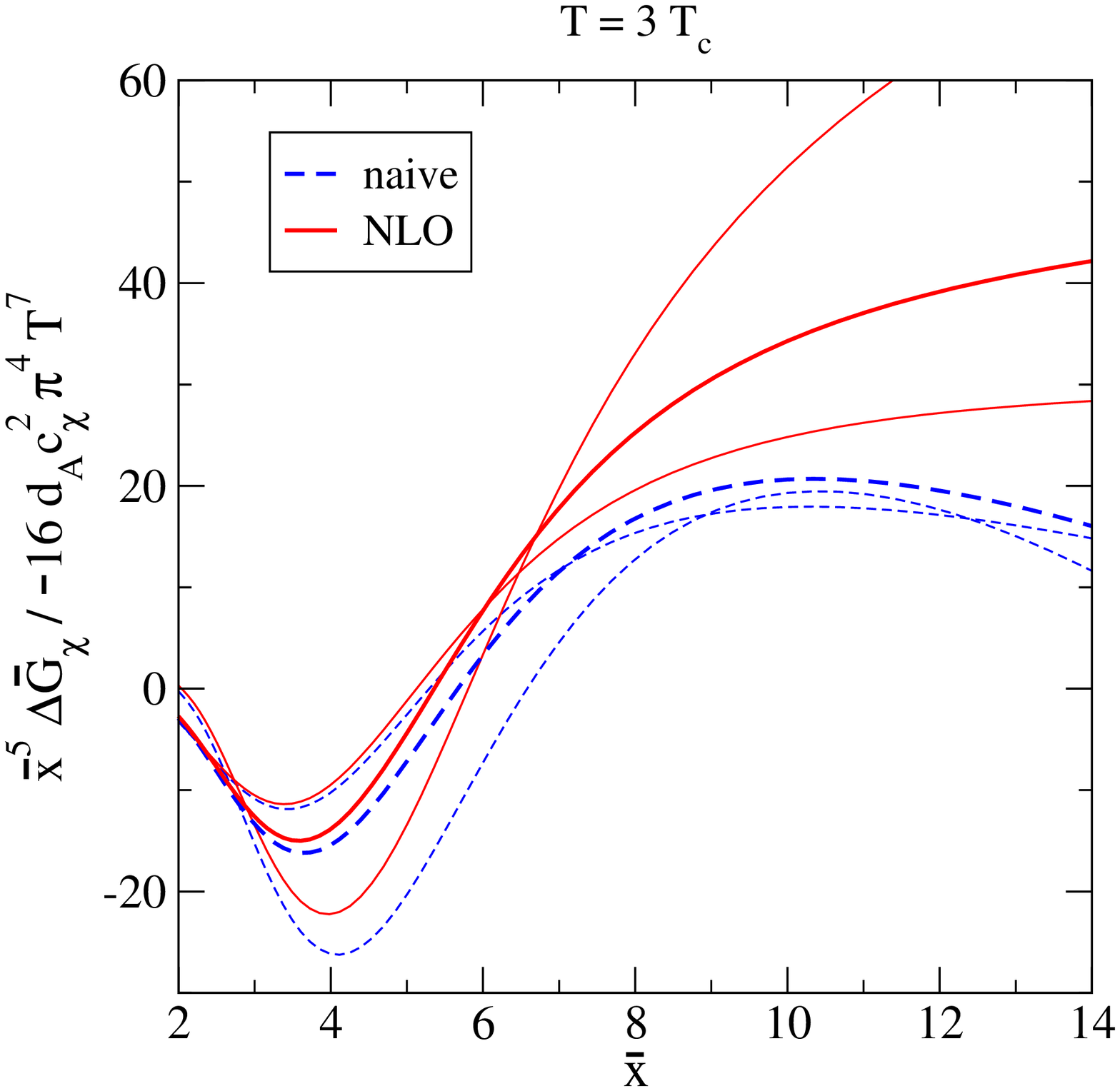}
}

\caption[a]{\small
The long-distance behaviours of the 
$\theta$ (left) and $\chi$ (right) correlators, multiplied by $\bar{x}^5$, 
at $T = 3\Tc$. The results are 
compared with the ``naive'' expressions which do not include 
the resummations of \eqs\nr{theta_resum}, \nr{chi_resum}.
Thick lines indicate 
the scale choice $\bar{\Lambda} = \bar{\Lambda}^\rmi{opt}$, thin 
lines variations by a factor of 2 around this scale. The ``naive''
results become negative at long distances, whereas the resummed
results remain positive and behave as $\sim 1/\bar{x}^5$ within 
our approximation. (This would turn into an exponential
decay by higher order EQCD corrections or by non-perturbative
MQCD effects.) 
}
\la{fig:rdep5}
\end{figure}
%%%%%%%%%%%%%%%%%%%%%%%%%%%%%%%%%%%%%%%%%%%%%%%%%%%%%%%%%%%%%%%%%%%%%%%%%%%

We start by inspecting the behaviour of the full 
correlators 
over a wide distance range. This leads to 
the plots shown in \figs\ref{fig:rdep1}, \ref{fig:rdep1_b} where we display 
\ba
 \frac{\bar G_\theta(x)}{4 d_A c_\theta^2\, \pi^4T^7}
     \quad\quad {\rm and} \quad\quad
 \frac{\bar G_\chi(x)}{-16 d_A c_\chi^2\, \pi^4T^7}
 \;, \la{for_display}
\ea
multiplied by $\bar{x}^6$ in order to increase the resolution, 
on a log-log scale at 
the temperatures $T=3\Tc$ ($\Tc \equiv 1.25 \Lambdamsbar$)
and $T=12\Tc$ 
for $\bar{\Lambda}=n\,\bar{\Lambda}^\rmi{opt}$, $n=\fr12,1,2$. 
A power-law behaviour can be clearly observed at short distances, 
whereas the dependence of the result on the renormalization 
scale as well as the magnitude of the NLO correction become
visible at long distances, indicating that 
the perturbative series becomes less trustworthy there. 

In order to observe more clearly the fine
structures of the correlators, we next consider the functions
\ba
 \frac{\Delta \bar G_\theta(x)} 
 {4 d_A c_\theta^2\, \pi^4T^7}
 &\equiv&
  \frac{\bar G_\theta(x)-\bar G^\rmi{``vac''}_\theta(x)}
  {4 d_A c_\theta^2\, \pi^4T^7}
  \;, \la{delta_theta} \\
 \frac{\Delta \bar G_\chi(x)}
 {-16 d_A c_\chi^2\, \pi^4T^7}
 &\equiv&
 \frac{\bar G_\chi(x)-\bar G^\rmi{``vac''}_\chi(x)}
 {-16 d_A c_\chi^2\, \pi^4T^7}
 \;. \la{delta_chi}
\ea
Here the vacuum-like parts are {\em defined} through the 
first lines of \eqs\nr{UVtheta}, \nr{UVchi} with 
a particular scale choice: 
\ba
 \frac{\bar G^\rmi{``vac''}_\theta(x)}{4 d_A c_\theta^2} 
 &\equiv &  
 \fr{g^4\pi^4 }{(2\pi {x})^7}
 \biggl\{480+\fr{g^2\Nc}{\pi^2} 
 \biggl[ 
   440\ln\(e^{\gammaE}\bar{\Lambda}x\) -\fr{824}{3}
 \biggr] \biggr\}_{ \bar{\Lambda} = n\, \bar{\Lambda}^\rmii{opt} } 
 \;, 
 \label{T0theta}\\
%%%%%
 \frac{\bar G^\rmi{``vac''}_\chi(x)}{-16 d_A c_\chi^2} 
 &\equiv &
 \fr{g^4\pi^4 }{(2\pi {x})^7}
 \biggl\{480+\fr{g^2\Nc}{\pi^2} 
 \biggl[ 
   440\ln\(e^{\gammaE}\bar{\Lambda}x\) -\fr{104}{3}
 \biggr] \biggr\}_{ \bar{\Lambda} = n\, \bar{\Lambda}^\rmii{opt} }
 \;. \label{T0chi}
\ea
Although it would be tempting to interpret these as genuine
vacuum results, this is not correct within our 
fixed-order computation: the genuine vacuum results
become non-perturbative at a distance $x \sim 0.1$~fm and this is 
reflected by the fact that with the scale choices of \eq\nr{Lamx}
the perturbative corrections grow out of control 
at $x\gsim 0.1\, \Lambdamsbar^{-1}$~\cite{lvv}. In contrast, 
in the full finite-temperature results of \eq\nr{for_display} 
a scale choice like that in \eq\nr{lambdaoptgen} 
is well-motivated, and at sufficiently high 
temperatures the result is perturbative 
(apart from MQCD contributions)
even at distances $x\sim$~fm.
The definitions of \eqs\nr{delta_theta}--\nr{T0chi} therefore only 
serve purposes of illustration.  
In any case, in \figs\ref{fig:rdep4}--\ref{fig:rdep5} 
these functions are plotted on various distance scales.

Considering first the UV limits of the correlators, 
in \fig\ref{fig:rdep4} we display our results 
for $\Delta \bar G_\theta$ and $\Delta \bar G_\chi$ 
multiplied by $\bar{x}^3$, with $\bar{x}$ ranging from 0 to 6, 
and compare the results with the UV-limits in 
\eqs\nr{UVtheta}, \nr{UVchi}. 
The results indicate that OPE-expanded results 
are applicable only in the range $\bar{x} \lsim 1$, 
and that there is a visible difference between the 
two channels in the regime $1\lsim \bar{x} \lsim 5$.

Next, motivated by the results of ref.~\cite{Iqbal:2009xz}, 
we inspect intermediate distances, $3\leq \bar{x} \leq 12$. 
The results are plotted in \fig\ref{fig:rdep3}. 
The results suggest
that the two channels are rather close to each other 
in the distance range $5 \lsim \bar{x} \lsim 10$.

Finally, \fig\ref{fig:rdep5} shows 
the long-distance behaviours of 
$\Delta \bar G_\theta$ and $\Delta \bar G_\chi$
multiplied by $\bar{x}^5$ (so that, 
within our approximation, they approach constants 
at large $\bar{x}$). We compare the curves to the result 
of a calculation where no resummation of the soft modes 
was carried out. We observe that 
the $- 1/\bar{x}^4$ behaviour of the unresummed result 
becomes fully visible only when $\bar{x}\gsim 10$.
Further higher order or non-perturbative corrections, 
which would eventually turn the behaviour in both 
channels into exponential decay~\cite{Arnold:1995bh}, 
can be assumed to become significant in the same range 
where resummation starts to dominate the behaviour, 
i.e.\ $\bar{x} \gsim 10$ at this temperature. 

%%%%%%%%%%%%%%%%%%%%%%%%%%%% SECTION %%%%%%%%%%%%%%%%%%%%%%%%%%%%%%%%%%%%%
%
\section{Conclusions}
\la{se:concl}

The purpose of this paper has been to ``interpolate'' 
between previous studies of the spatial correlation functions 
related to the operators $\theta\propto \tr[F_{\mu\nu} F_{\mu\nu}]$ and 
$\chi\propto\tr[F_{\mu\nu} \tilde F_{\mu\nu}]$ within pure SU(3) gauge 
theory at a finite temperature $T \sim$ a few $\times\; \Tc$.  
We find that the ``ultraviolet limit'', 
studied in ref.~\cite{lvv}
(following refs.~\cite{Iqbal:2009xz,sch} and other recent works), 
in which the correlators 
are computed with Operator Product Expansion type methods, 
works reasonably only at distances 
$2\pi T x \lsim 1$ (cf.\ \fig\ref{fig:rdep4}).  
At the same time the ``infrared limit'', studied 
in ref.~\cite{Arnold:1995bh} and many subsequent works, 
starts to govern the behaviour at distances 
$2\pi T x \gsim 10$ (cf.\ \fig\ref{fig:rdep5}).
In between, neither of these simplified approaches 
is viable, and our results represent the most precise
weak-coupling expressions available to date. Whether 
they themselves are quantitatively accurate remains to be seen, 
once lattice measurements for the 
very same quantities, 
following the lines that have recently been published 
in ref.~\cite{Iqbal:2009xz}, become available.

Among the more detailed motivations of our study was 
the question, to what extent do the two correlators differ from each 
other? While we have indeed found some disparities between the 
two channels, in the direction that all features are more 
pronounced in the correlator 
corresponding to $\tr[F_{\mu\nu} \tilde F_{\mu\nu}]$, 
the difference is not as striking as that seen in the lattice study of 
ref.~\cite{Iqbal:2009xz}. Several possible reasons come to mind. 
First of all, the leading ``conformality-breaking'' OPE contribution
proportional to $(e-3p)(T)$, showing a clear distinction 
between the two channels at short distances, corresponds 
actually to the 3-loop order in terms of Feynman graphs
(in ref.~\cite{lvv} it could be ``guessed'' from 
the 2-loop computation carried out, by making use of 
renormalization scale invariance in connection with the general 
factorization philosophy inherent to OPE). This might suggest 
that 3-loop contributions are important insofar as the difference
between the two channels is concerned. Second, it must be 
kept in mind that the lattice measurements 
in ref.~\cite{Iqbal:2009xz}
concerned strictly
local correlators, whereas ours have been averaged over the 
Euclidean time coordinate. 
This has an important effect at least at distances $x\lsim \beta$, 
i.e.\ $2\pi T x \lsim$~a few.
It will be interesting to repeat
the comparison if lattice results for time-averaged correlators 
become available. 
Third, it should be stressed than in ref.~\cite{Iqbal:2009xz}
the genuine vacuum parts were subtracted from the thermal correlators. 
Unfortunately, this makes the difference sensitive to 
confinement physics; the features seen may {\em not} be
a reflection of perturbative thermal physics alone. It would be 
useful to carry out future comparisons {\em without} such 
a subtraction, i.e.\ directly 
with \figs\ref{fig:rdep1}, \ref{fig:rdep1_b}.

Let us remark that the ``thermal'' parts of the correlators, 
as well as the difference between the two ``thermal'' parts, 
show a remarkably rich $x$-dependence, suggesting that
a high resolution is needed for disentangling all relevant features. 
As can be observed in \figs\ref{fig:rdep4}--\ref{fig:rdep5}, 
the ``thermal'' parts change sign twice, before finally showing a positive
correlation at long distances. At the same time, the difference
between the two ``thermal'' parts, proportional to the 
thermal part of the function 
$\phi_\rmi{LO}$ (cf.\ \eq\nr{reschi} and \fig\ref{fig:basic_a}), 
shows the same two sign changes at short and intermediate 
distances, but would have a third one in store for long 
distances: at the ``realistic'' temperatures 
considered, the screening mass corresponding to 
$\tr[F_{\mu\nu} \tilde F_{\mu\nu}]$ is markedly larger
than that corresponding to $\tr[F_{\mu\nu} F_{\mu\nu}]$
(cf.\ refs.~\cite{Hart:2000ha,Laine:2009dh}), implying
that at very large distances the correlator $\bar{G}_\chi$ 
would be smaller in magnitude than $\bar{G}_\theta$.

We end by recalling the relation of our work to heavy quarkonium
physics and simultaneously the physical reason for considering
time-averaged correlators as we have done in the present study.  
Whereas it was traditionally thought that heavy quarkonium melting
is primarily a consequence of colour-electric Debye screening
at long distances, $x \sim 1/g T$, it has been suggested in recent
years that at least in the bottomonium case, where weak-coupling
techniques are expected to be semi-quantitatively applicable, 
the ``melting'' of quarkonium resonances takes place already when 
the Bohr radius of the bound state, $r$, is parametrically in 
the range $1/\pi T \ll r \ll 1/g T$~\cite{Burnier:2007qm}--\cite{dw}.
In this situation the dominant mechanism for melting is not
Debye screening but Landau damping or, more physically,  
the decoherence of the quantum mechanical bound state, mathematically
represented by an imaginary part in the real-time static 
potential felt by a heavy quark-antiquark pair propagating
in Minkowski time~\cite{Laine:2006ns} 
(see also refs.~\cite{Laine:2007qy}--\cite{rhs}). Moreover, 
it can be argued that it is interesting to study quarkonium
states even before they melt, whence the regime $x \lsim 1/\pi T$ 
becomes relevant as well~\cite{Brambilla:2010vq}. 
With such motivations, the ``singlet free energy'' 
as determined from lattice simulations~\cite{q_singlet} 
and traditionally used in potential models has recently been 
compared with perturbation theory at intermediate distances $x\sim 1/\pi T$, 
and a surprising agreement between the two 
was found~\cite{Burnier:2009bk}.\footnote{%
 A similar if inequivalent observable, the correlator of
 two ``traced'' Polyakov loops, sometimes referred to as 
 (exponent of minus) the 
 ``colour-averaged potential'', was computed perturbatively 
 in ref.~\cite{Brambilla:2010xn}, but no numerical evaluation
 or comparison with lattice data was presented. 
 } 
The singlet free energy consists precisely of a correlator
of two time-integrated objects. It is 
for this reason that we suspect time-averaged correlators to be 
of more direct physical relevance than strictly local ones. 
By learning about the reliability of the weak-coupling expansion
for observables of this type in the intermediate distance range, 
it will hopefully become possible in the end to relate  
perturbative determinations of heavy quarkonium 
spectral functions (ref.~\cite{Burnier:2008ia} and references therein), 
which use the real-time static potential as input,  
to on-going lattice studies such as those 
in refs.~\cite{Aarts:2010ek,Ding:2010yz}, 
thereby learning about the systematic errors on both sides.

%%%%%%%%%%%%%%%%%%%%%%%%% SECTION %%%%%%%%%%%%%%%%%%%%%%%%%%%%%%%%%%%%%
%
\section*{Acknowledgements}

We thank K.~Kajantie and H.B.~Meyer for useful discussions. 
M.L.\ was partly supported by 
the BMBF under project
{\em Heavy Quarks as a Bridge between
     Heavy Ion Collisions and QCD}; 
M.V.\ was supported by the Academy of Finland, 
contract no.\ 128792, and 
A.V.\ by the Sofja Kovalevskaja program 
of the Alexander von Humboldt foundation.

%%%%%%%%%%%%%%%%%%%%%%% APPENDIX %%%%%%%%%%%%%%%%%%%%%%%%%%%%%%%%%%%
%
\appendix
\renewcommand{\thesection}{Appendix~\Alph{section}}
\renewcommand{\thesubsection}{\Alph{section}.\arabic{subsection}}
\renewcommand{\theequation}{\Alph{section}.\arabic{equation}}

%%%%%%%%%%%%%%%%%%%%%%%%%%%%%% SECTION %%%%%%%%%%%%%%%%%%%%%%%%%%%%%%%%%
%
%\section{Appendix}

%%%%%%%%%%%%%%%%%%%%%%%%%%%%%%%%%%%%%%%%%%%%%%%%%%%%%%%%%%%%%%%%%%%%%%%%%%%
%

\end{document}